\documentclass[a4paper,11pt]{article}
\pdfoutput=1 

\usepackage{jheppub} 

\usepackage[english]{babel}
\usepackage{bm,amsmath,amssymb,slashed,graphicx,%
            enumerate,alltt,xspace,multirow,xcolor,mathrsfs}
\usepackage{fancyvrb}
\usepackage{cancel}

\usepackage[normalem]{ulem}

\usepackage{subcaption}

\RequirePackage{braket}
\RequirePackage{graphicx}

\RequirePackage[numbers,sort&compress]{natbib}

\makeatletter
\g@addto@macro\bfseries{\boldmath}
\makeatother

\definecolor{labelkey}{rgb}{0,0.5,0.0}

\usepackage{listings}
\definecolor{darkgreen}{rgb}{0,0.4,0}

\newcommand\mathd{\mathrm{d}}

\newcommand{\as}{\alpha_s}
\newcommand{\aem}{\alpha}

\newcommand{\muF}{\mu_{\scriptscriptstyle \rm F}}

\newcommand{\pt}{p_{\rm \scriptscriptstyle T}}

\newcommand\HT{HT}

\newcommand\elgam{$\ell/\gamma$}
\newcommand\elgaminit{\elgam{}-initiated}
\newcommand\el{$\ell$}
\newcommand\elinit{\el{}-initiated}
\newcommand\gam{$\gamma$}
\newcommand\gaminit{\gam{}-initiated}

\newcommand\yl{y_{\ell}}
\newcommand\ggam{g_{\gamma}}

\newcommand{\LUXlep}{{\tt LUXLep}}
\newcommand{\LUX}{{\tt LUX}}

\newcommand{\hoppet}{{\tt Hoppet}}
\newcommand{\LHAPDF}{{\tt LHAPDF}}
\newcommand{\POWHEGBOX}{{\tt POWHEG-BOX}}
\newcommand{\MADGRAPH}{{\tt Madgraph}}

\newcommand{\POWHEG}{{\tt POWHEG}}
\newcommand{\Herwig}{{\tt Herwig7}}
\newcommand{\Pythia}{{\tt Pythia8}}

\newcommand\LHENLO{LHE}

\newcommand{\tmop}[1]{\ensuremath{\operatorname{#1}}}

\newenvironment{enumerateroman}{\begin{enumerate}[i.] }{\end{enumerate}}
\newenvironment{itemizedot}{\begin{itemize} }{\end{itemize}}

\title{Photon and Leptons induced processes at the LHC}
\preprint{
  \begin{flushright}
    ZU-TH 44/21\\
    MPP-2021-166
  \end{flushright}
}

\author[a]{Luca Buonocore,}
\author[b]{Paolo Nason,}
\author[c]{Francesco Tramontano,}
\author[d]{Giulia Zanderighi}

\emailAdd{lbuono@physik.uzh.ch}
\emailAdd{paolo.nason@mib.infn.it}
\emailAdd{francesco.tramontano@cern.ch}
\emailAdd{zanderi@mpp.mpg.de}

\affiliation[a]{University of Zurich, Winterthurerstrasse  190, 8057 Zurich, Switzerland}
\affiliation[b]{Universit\`a di Milano-Bicocca and INFN, Sezione di
  Milano-Bicocca, Piazza della Scienza 3,20126 Milano, Italy}
\affiliation[c]{Universit\`a di Napoli and INFN, Sezione di
  Napoli, Complesso Universitario di Monte Sant'Angelo,
  Via Cinthia 21, 80126 Napoli, Italy} 
\affiliation[d]{Max-Planck-Institut f\"ur Physik, F\"ohringer Ring 6,
  80805 M\"unchen, Germany}

\date{Received: date / Accepted: \today}

\abstract{We study a few basic photon- and lepton-initiated processes
  at the LHC which can be computed using the recently developed photon
  and lepton parton densities.  First, we consider the production of a
  massive scalar particle initiated by lepton-antilepton annihilation and
  photon-photon fusion as representative examples of searches of
  exotic particles.  Then we study lepton-lepton scattering, since
  this Standard-Model process may be observable at the LHC.  We
  examine these processes at leading and next-to-leading order and,
  using the \POWHEG{} method, we match our calculations to parton
  shower programs that implement the required lepton or photon
  initial-states.  We assess the typical size of cross-sections and
  their uncertainties and discuss the preferred choices for the
  factorization scale.
  These processes can also be computed starting directly from the
  lepto-production hadronic tensor, leading to a result where some
  collinear-enhanced QED corrections are missing, but all strong
  corrections are included. Thus, we are in the unique position to
  perform a comparison of results obtained via the factorization
  approach to a calculation that does not have strong corrections.
  This is particularly relevant in the case of lepton-scattering, that
  is more abundant at lower energies where it is affected by
  larger strong corrections. We thus compute this process
  also with the hadronic-tensor method, and compare the results with those
  obtained with \POWHEG{}.
Finally, for some lepton-lepton scattering processes, we compare the
size of the signal to the main quark-induced background, which is
double Drell-Yan production, and outline a preliminary search strategy
to enhance the signal to background ratio.}


\begin{document}


\maketitle


\newcommand{\citere}[1]{Ref.\,\cite{#1}}
\newcommand{\citeres}[1]{Refs.\,\cite{#1}}

\section{Introduction}
\label{sec:intro}
In recent years, precise determinations of the proton Parton-Density
Functions (PDFs) of the photon~\cite{Manohar:2016nzj, Manohar:2017eqh}
(from here on the \LUX{} papers) and leptons~\cite{Buonocore:2020nai}
(from here on the \LUXlep{} paper) have appeared.  These determinations
(inspired by~\citere{Drees:1988pp}) were based upon the comparison of
the computation of a reference photon- or lepton-initiated process
(from here on, \elgaminit{} process) in the collinear factorization
approach, and in terms of the proton electromagnetic hadronic tensor
\begin{equation}
  W_{\mu\nu}(p,q)=\frac{1}{4\pi}\int \mathd^4z e^{i\, q\cdot z}\langle
  p|[J_{\mu}(z),J_\nu(0)]|p\rangle,
\end{equation}
where $p$ denotes the proton momentum, the bra and ket states refer to
a single proton state, and we imply an average over its spin. The
hadronic tensor can be expressed in terms of the electroproduction
structure functions and of the proton electromagnetic form factors. We
will denote this method for the computation of \elgaminit{} processes
in hadronic collisions as the \HT{} (for Hadronic Tensor) method.

The calculation in the factorization approach is analogous to the
Weizs\"acker-Williams approximation in charged particle scattering. It
starts with the simple leading order (LO) approach, where one only
needs the Born matrix elements for the \elgaminit{} process, and is
easily extended to next-to-leading order (NLO) and even to higher
orders, depending upon the accuracy of the available PDF.
It can benefit from available packages for the
computation of NLO corrections, and also for implementing NLO
calculations interfaced to parton showers (NLO+PS from here on).

So far, most calculations of \elgaminit{} processes using the
factorization approach have been performed at the LO level, with the
noticeable exception of~\citere{Greljo:2020tgv}, where the full NLO
corrections to the resonant $s$-channel leptoquark production was
presented.  On the other hand, the photon and lepton PDF
determinations presently available have NLO accuracy. Thus it would be
desirable to fully exploit them using NLO calculations.  As a related
point, while we have considerable experience regarding the relative
size of NLO corrections in processes initiated by quarks and gluons,
little is known for \elgaminit{} processes in hadronic collisions.  In
the present paper we discuss some of these processes, compute them
at LO and NLO, interface the calculations to parton shower
generators according to the \POWHEG{}
method~\cite{Nason:2004rx,Frixione:2007vw,Alioli:2010xd}, and study
their scale uncertainties.

We remark that NLO corrections in \elgaminit{} processes arise in a
rather peculiar way. Rather than adding gluons to the process, as
perhaps one would naively expect, corrections that are formally of NLO
importance arise when the incoming photons or leptons are resolved as
radiation from a parton that has higher abundance in the proton. For
example, if we have an incoming photon, we also have a diagram where
the photon arises from quark radiation. This diagram leads to a contribution
proportional to a quark density times one power of the electromagnetic coupling
$\aem$. Being a subtracted NLO contribution it does not have any
logarithmic enhancement. On the other hand, the corresponding Born
process multiplies a photon density, that is of the order of a quark
density times one power of the electromagnetic coupling, times a
collinear logarithm $L=\log(Q/\Lambda)$, where $\Lambda$ is a typical
hadronic scale. Such logarithm is parametrically of the same size of
the inverse of the strong coupling constant $\as(Q)$, so that indeed
the NLO correction is suppressed by a power of $\as(Q)$ with respect
to the Born term.

An interesting aspect of \elgaminit{} processes is the possibility to
search for exotic signals, typically resonances, that may preferably
or exclusively couple to leptons or photons. We thus consider the
production of a scalar via the processes $l^+ l^- \to \phi$ and
$\gamma\gamma \to \phi$, study their NLO corrections and implement
corresponding NLO+PS generators using the \POWHEG{} method.  In the
case of the $\gamma\gamma \to \phi$ process, we can interface our
\POWHEG{} generator to both \Pythia{}~\cite{Sjostrand:2014zea} and
\Herwig{}~\cite{Bellm:2019zci}. Incoming leptons are not handled at
the moment by \Pythia{}, while there is a preliminary implementation
in \Herwig{}, so, in the lepton case, we restrict ourselves to
consider the latter.  We can thus assess the impact of NLO corrections
both in the fixed order calculation, and in the full NLO+PS
implementation.  It is also interesting to compare the full NLO+PS
results to LO+PS ones, since often beyond the Standard Model (BSM)
studies are carried out at this level.

In the \LUXlep{} paper, it was pointed out that some Standard Model (SM)
lepton-scattering processes leading to rare final states, like leptons
pairs of different flavours or same sign, may actually be detectable
at the LHC, and some LO cross sections for these processes were given.
Here we examine this problem in more depth, by including NLO
corrections and implementing an NLO+PS generator interfaced to
\Herwig{}.  The same processes can also be computed directly adopting
the \HT{} approach.\footnote{A result in this framework has been given
  in Ref.~\cite{Harland-Lang:2021zvr}.  In the present work we include
  contributions that were omitted in~\cite{Harland-Lang:2021zvr}, as
  will be discussed in the introduction of section~\ref{sec:lepscat}.}
We are thus in a position to compare the full NLO calculation with the
\HT{} result, thereby validating the NLO+PS approach and also the
Monte Carlo implementation of lepton initiated processes.

The factorization approach is based upon an extraction of the \elgam{}
parton densities from the result of an \HT{} calculation, neglecting
terms that are beyond the required accuracy (i.e. typically beyond
NLO). Thus, the \HT{} method could be in principle more precise. 
However it is more cumbersome to implement and, in the case of
incoming leptons, it neglects electromagnetic corrections that are
included with the factorization method, namely radiation of collinear
photons from the leptons entering the scattering process.  Often when
considering mixed electromagnetic and strong corrections, two powers
of $\as$ are considered to be equivalent to a single power of the
electromagnetic coupling constant. According to this counting, initial
state electromagnetic radiation from incoming leptons should count as
$\aem L \sim \aem/\as(Q)$, i.e. as NLO contributions.  We are however in a
position to assess the size of these effects, by generating a lepton
PDF set where initial-state radiation from leptons has been turned
off, and thus we can carry out a meaningful comparison of the two
methods.

The paper is divided as follows. We start by discussing in
Sec.~\ref{sec:simple} the production of a scalar particle either from
two initial state photons, or from a lepton-antilepton pair.  We discuss
the size of the cross-sections and their scale uncertainties in
Sec.~\ref{sec:scalartot}, and present their differential distributions
in Sec.~\ref{sec:scalarfromphotons} and~\ref{sec:scalarfromleptons}.
In Sec.~\ref{sec:lepscat} we discuss lepton scattering processes. We
restrict ourselves to those final states that do not receive contributions 
from the overwhelming
Drell Yan background at leading order. We discuss the corresponding
production cross-sections with their uncertainty bands at LO and NLO
in Sec.~\ref{sec:lepscatNLO}. We then discuss in Sec.~\ref{sec:cuts}
the cuts required to suppress large SM backgrounds and their impact on
the signal. We also present a comparison to predictions obtained with
the \HT{} method.
We give our conclusions in
Sec.~\ref{sec:conclu}. Appendix~\ref{app:HT} provides details about
the calculation in the \HT{} approach, including basic formulae
(\ref{app:basic}) and the phase space
(\ref{app:PS}).
In Appendix~\ref{app:sfsplitmass} we describe how we include
lepton mass effects to cure the final-state collinear divergences
arising from photon to lepton splitting contributions in the NLO
calculation.

\section{$\gamma \gamma \to \phi$ and $l^+ l^- \to \phi$}
\label{sec:simple}

In order to better understand the impact of the NLO QCD corrections to
\elgaminit{} processes we focus on a simple case, namely the
production of a neutral scalar resonance which couples only either to
photons or to leptons.  In particular, the objective of this section
is to assess the size of the NLO corrections and their dependence upon
the scale choice. Furthermore, we compare selected kinematic
distributions at LO and NLO matched with parton shower using the
\POWHEG{} method.

\noindent We describe the interaction among photons and the scalar
resonance through the effective Lagrangian
\begin{equation}
  \mathcal{L}_{\rm eff}^{(\gamma)} =
  -\frac{1}{4}\frac{\ggam}{M_\phi}\phi
  \mathcal{F}^{\mu\nu}\mathcal{F}_{\mu\nu}\,,
\end{equation}
where $\mathcal{F}^{\mu\nu}$ is the electromagnetic field strength
tensor, $M_{\phi}$ is the mass of the scalar resonance and
$g_{\gamma}$ is a dimensionless effective coupling. For the leptons,
we consider a simple effective Yukawa interaction
\begin{equation}
  \mathcal{L}_{\rm eff}^{(\ell)} = -i \yl \phi {\bar \ell} \ell.
\end{equation}
Our processes proceed at LO via the partonic sub-processes $\gamma
\gamma \to \phi$, in one case, and $l^\pm l^\mp \to \phi$, in the
other.  According to the discussion given in the introduction, they
receive NLO QCD corrections from the partonic processes $q\gamma \to q \phi$
(and the corresponding crossed one $q {\bar q}\to \gamma \phi$, which is finite) and $l^{\pm} \gamma \to l^\pm \phi$ respectively.
In fact, while on one hand the latter processes
are down by a power of $\alpha$, they are enhanced by the ratio
$f_q/f_\gamma \sim 1/ (\alpha L)$ and $f_\gamma/f_l \sim \alpha L /
(\alpha L)^2 = 1/ (\alpha L)$ respectively, resulting in an order $1/L
\sim \alpha_s$ correction (for a more detailed discussion see the
\LUXlep{} paper).
\subsection{Total cross sections and scale dependence}
\label{sec:scalartot}
\begin{figure}[htb]
  \centering
  \includegraphics[width=\textwidth]{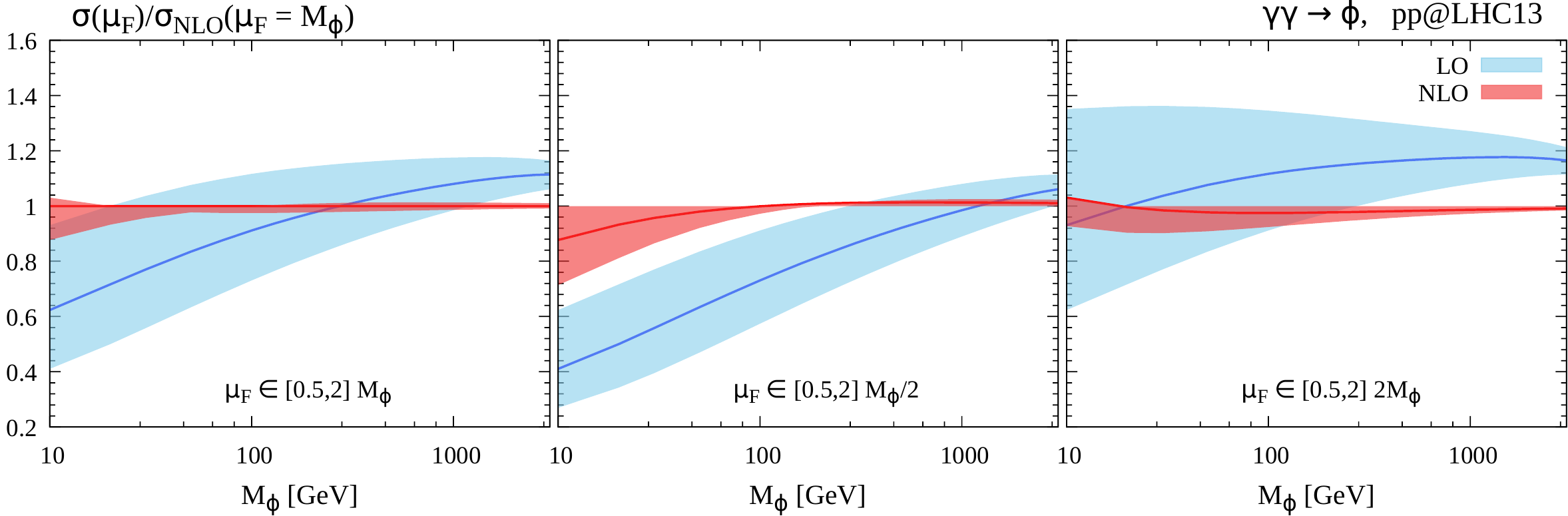}
  \caption{Comparison of LO and NLO cross sections for the
    photon-induced process in proton-proton collisions at
    $\sqrt{s}=13\,$TeV, normalised to the reference NLO cross
    section evaluated at $\mu_F=M_\phi$, as a function of the mass of
    the scalar particle. The three panels correspond to the different
    choices of the central scale $\mu_0=(1,1/2,2)M_\phi$. The
    uncertainty bands are associated to the 3-point variation of the
    factorization scale in the nominal range $ 1/2 < \mu_F/\mu_0 <
    2$.}\label{fig:scalevar-gagaPhi}
\end{figure}

In order to understand the relative impact of the NLO corrections and
the remaining uncertainties associated to missing higher orders, we
show in Figures~\ref{fig:scalevar-gagaPhi},
\begin{figure}[htb]
  \centering
  \includegraphics[width=\textwidth]{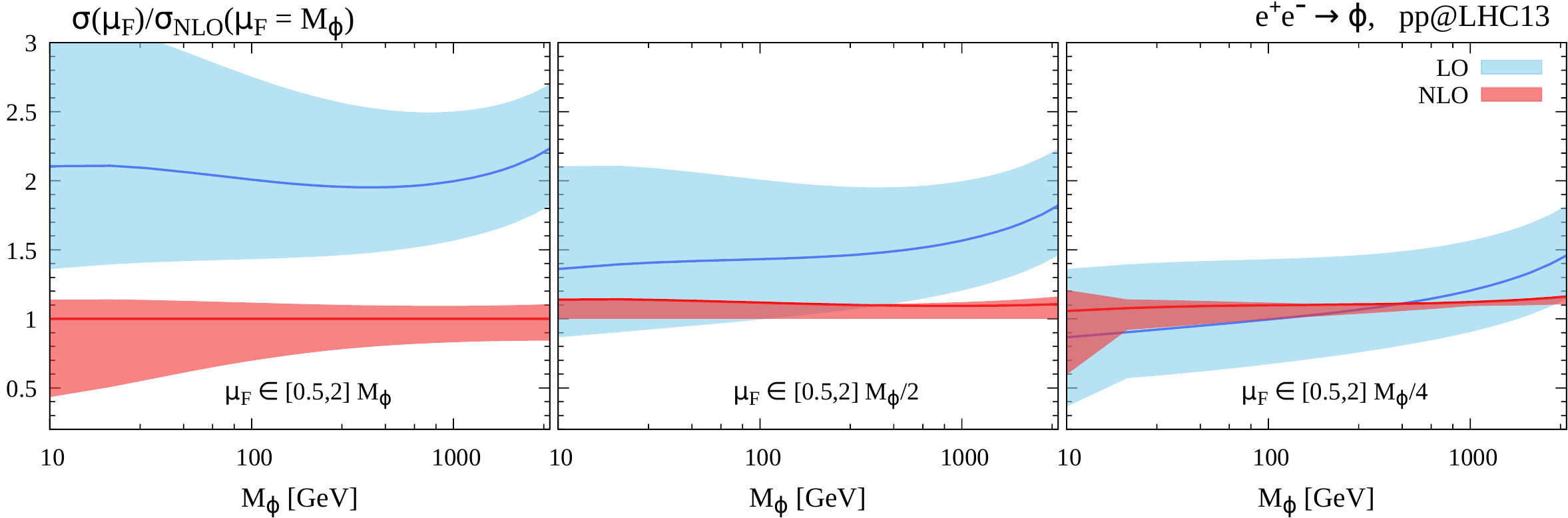}
  \caption{Comparison of LO and NLO cross sections for the
    lepton-induced process in proton-proton collisions at
    $\sqrt{s}=13\,$TeV, normalised to the reference NLO cross
    section evaluated at $\mu_F=M_\phi$, as a function of the mass of
    the scalar particle. The three panels correspond to the different
    choices of the central scale $\mu_0=(1,1/2,1/4)M_\phi$. The
    uncertainty bands are associated to the 3-point variation of the
    factorization scale in the nominal range $ 1/2 < \mu_F/\mu_0 <
    2$.}\label{fig:scalevar-eePhi}
\end{figure}
the LO and NLO cross sections for the
\gaminit{} process in proton-proton collisions at
$\sqrt{s}=13\,$TeV, as a function of the scalar mass and for
different choices of the central factorization scale, normalised to
the NLO one computed at a fixed reference scale $\mu_F=M_\phi$.
The analogous plot for the \elinit{} process is reported in Fig.~\ref{fig:scalevar-eePhi}.
Here and in the following, to obtain our results
we have used the PDF set LUXlep-NNPDF31\_nlo\_as\_0118\_luxqed of
Ref.~\cite{Buonocore:2020nai}, as implemented in
\LHAPDF{}~\cite{Buckley:2014ana}.
The bands display the associated scale uncertainty
corresponding to the 3-point variation of the factorization scale, $
\muF = (1/2 , 1, 2) \mu_0 $, $\mu_0$ being the central scale.

First,
we observe that the LO result is affected by sizeable scale
uncertainties which display a qualitatively similar behavior in all
cases: they range from about $50\%$ at $M_\phi=10\,$GeV to about 5\%
(20\%) at $M_\phi=2\,$TeV for the photon (lepton) induced process. The
scale uncertainties shrink considerably going from LO to NLO, being of
few percent (10\%) for the photon (lepton) induced process, with a
mild dependence upon $M_\phi$. Furthermore, we observe that the
different scale-range choices give rather consistent results at
NLO. On the other hand, we see that the LO predictions are only
indicative of the cross section within a factor of order one, as is
typical for LO predictions for collider processes.  We find that the
choice of the mass of the scalar particle as central scale is
appropriate for the photon induced process, leading to K-factors
around unity in the whole mass range. The situation is different for
the lepton induced process, where this choice leads to large and
negative corrections (K-factors~$\sim 1/2$). This behavior has also
been observed in the leptoquark production
case~\cite{Greljo:2020tgv}. We also observe that in this case for the
central scale $\mu_0=M_{\phi}$ the LO and NLO bands do not overlap. An
optimal choice is provided by a smaller central scale, namely
$\mu_0=M_\phi/4$.  In Tab.~\ref{tab:xsecs_gagaphi_eephi}, we report a
collection of LO and NLO cross sections at LHC at 13 TeV for different
masses of the scalar resonance, adopting the above optimal choice for
the scales, and factoring out the corresponding $\ggam$ and $\yl$ couplings. We observe that
photon induced cross sections at a given mass are about a factor
$10^4$ larger than the corresponding lepton induced ones. This
reflects directly the relative abundance of photons and leptons in the
proton, since the two LO matrix elements have a constant ratio of
order 1.
\begin{table}
  \centering
  \resizebox{\columnwidth}{!}{
  \begin{tabular}{|c||c|c||c|c|}
    \hline
    \rule{0pt}{1\normalbaselineskip}
    $M_\phi\,$[GeV] & $\sigma^{(\gamma)}_{\rm LO}/\ggam^2\,$[pb] & $\sigma^{(\gamma)}_{\rm NLO}/\ggam^2\,$[pb] & $\sigma^{(\ell)}_{\rm LO}/\yl^2\,$[pb] & $\sigma^{(\ell)}_{\rm NLO}/\yl^2\,$[pb] \\ [0.15cm]
    \hline
    \rule{0pt}{1\normalbaselineskip}
    \hspace{-0.1cm}$10$ & $5.77^{+2.84}_{-1.97}$ $\cdot 10^{4}$  & $9.26^{+0.28}_{-1.13}$ $\cdot 10^{4}$ & $3.67^{+2.10}_{-2.13}$  & $4.48^{+0.64}_{-1.95}$  \\ [0.15cm]
    $50$   & $1.43^{+0.41}_{-0.34}$ $\cdot 10^{3}$   & $1.71_{-0.04}$        $\cdot 10^{3}$  & $1.07^{+0.52}_{-0.37}$ $\cdot 10^{-1}$  & $1.22^{+0.04}_{-0.14}$ $\cdot 10^{-1}$ \\ [0.15cm]
    $100$  & $2.26^{+0.51}_{-0.45}$ $\cdot 10^{2}$   & $2.48_{-0.06}$        $\cdot 10^{2}$  & $1.90^{+0.84}_{-0.62}$ $\cdot 10^{-2}$  & $2.09^{+0.03}_{-0.19}$ $\cdot 10^{-2}$ \\ [0.15cm]
    $250$  & $1.53^{+0.24}_{-0.23}$ $\cdot 10^{1}$   & $1.54^{+0.02}_{-0.03}$ $\cdot 10^{1}$  & $1.47^{+0.55}_{-0.43}$ $\cdot 10^{-3}$  & $1.53_{-0.09}$ $\cdot 10^{-3}$ \\ [0.15cm]
    $500$  & $1.57^{+0.19}_{-0.18}$    & $1.50^{+0.02}_{-0.03}$   & $1.56^{+0.53}_{-0.42}$ $\cdot 10^{-4}$  & $1.55_{-0.07}$ $\cdot 10^{-4}$ \\ [0.15cm]
    $1000$ & $1.17^{+0.10}_{-0.10}$ $\cdot 10^{-1}$  & $1.09^{+0.01}_{-0.01}$ $\cdot 10^{-1}$ & $1.11^{+0.33}_{-0.28}$ $\cdot 10^{-5}$  & $1.04_{-0.03}$ $\cdot 10^{-5}$ \\ [0.15cm]
    $1500$ & $2.06^{+0.15}_{-0.15}$ $\cdot 10^{-2}$  & $1.88^{+0.02}_{-0.02}$ $\cdot 10^{-2}$ & $1.79^{+0.50}_{-0.42}$ $\cdot 10^{-6}$  & $1.59_{-0.05}$ $\cdot 10^{-6}$ \\ [0.15cm]
    $2000$ & $5.14^{+0.31}_{-0.32}$ $\cdot 10^{-3}$  & $4.64^{+0.05}_{-0.05}$ $\cdot 10^{-3}$ & $4.04^{+1.08}_{-0.92}$ $\cdot 10^{-7}$  & $3.45_{-0.13}$ $\cdot 10^{-7}$ \\ [0.15cm]
    $3000$ & $5.22^{+0.24}_{-0.25}$ $\cdot 10^{-4}$  & $4.69^{+0.05}_{-0.04}$ $\cdot 10^{-4}$ & $3.33^{+0.82}_{-0.71}$ $\cdot 10^{-8}$  & $2.65^{+0.02}_{-0.13}$ $\cdot 10^{-8}$ \\ [0.15cm]
    \hline
  \end{tabular}
  }\caption{LO and NLO cross sections for a heavy scalar resonance
    production in \elgaminit{} processes at the LHC at 13 TeV as a
    function of the mass $M_{\phi}$. The uncertainties are obtained by
    performing the 3-point variation around the optimal scale choices
    $\muF=M_{\phi}$ for the $\gamma$-induced and $\muF=M_{\phi}/4$ for
    the $\ell$-induced process, as discussed in the
    text. } \label{tab:xsecs_gagaphi_eephi}
\end{table}

\subsection{Differential distributions for $\gamma \gamma \to \phi$}
\label{sec:scalarfromphotons}
We now examine differential distributions for the photon-induced
process at the level of full NLO+PS generation, that we computed using
the \POWHEG{} framework interfaced with Pythia 8.245, and Herwig 7.2
(both these generators implement processes initiated by
photons~\cite{Richardson:2010gz}).  In our analysis we do not include
multi-parton interactions, while we let the events shower and
hadronize. We use \Pythia{} with default parameters.  As for
\Herwig{}, we needed to set it up with customized options\footnote{We
  acknowledge Silvia Ravasio Ferrasio for the support in these
  modifications.}  in order to obtain a reasonable behavior in the
region of small transverse momentum of the scalar particle.  In
particular, we force the assignment of an intrinsic transverse momentum,
whose
default value is $2.2\,$GeV, also to colourless particles, which was
not the case with the default \Herwig{}. Another important aspect,
especially relevant when performing the matching at NLO, concerns the
recoil scheme (see for example Ref.~\cite{Cabouat:2017rzi}).  Most of
the NLO events are $q\gamma \to q \phi$. We remind that \Pythia{}
adopts by default a global recoil scheme for initial state radiation
(ISR), that corresponds to an initial-initial (II) dipole setup and
affects all final state particles. On the other hand, \Herwig{}
implements a strict dipole approach, that allows for both II and
initial-final (IF) recoils.  We observe that most of the time,
\Herwig{} treats the initial and final quark as an IF dipole,
enforcing a local recoil that does not change the transverse momentum
of the scalar. This leads to some distortions in the spectrum of the
resonance at small $p_T$. For these reasons, we force a global recoil
scheme also in \Herwig{}.

\begin{figure}[htb]
  \resizebox{7cm}{!}{\includegraphics{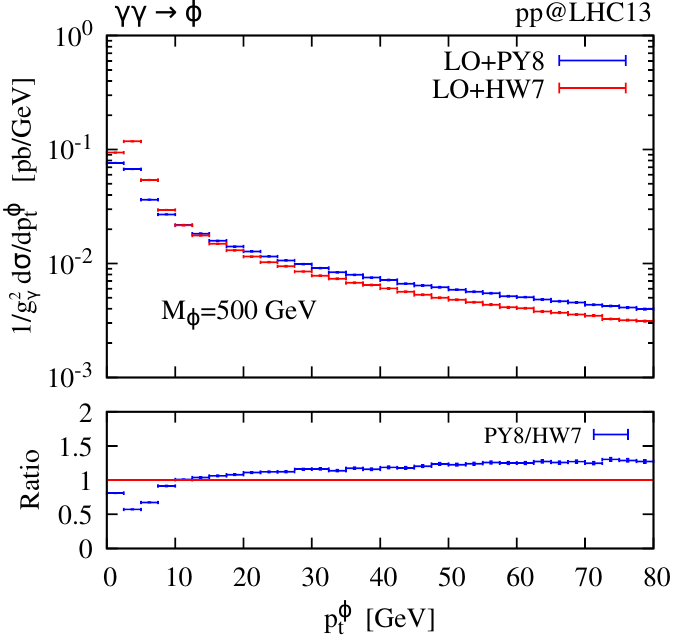}}
  \resizebox{7cm}{!}{\includegraphics{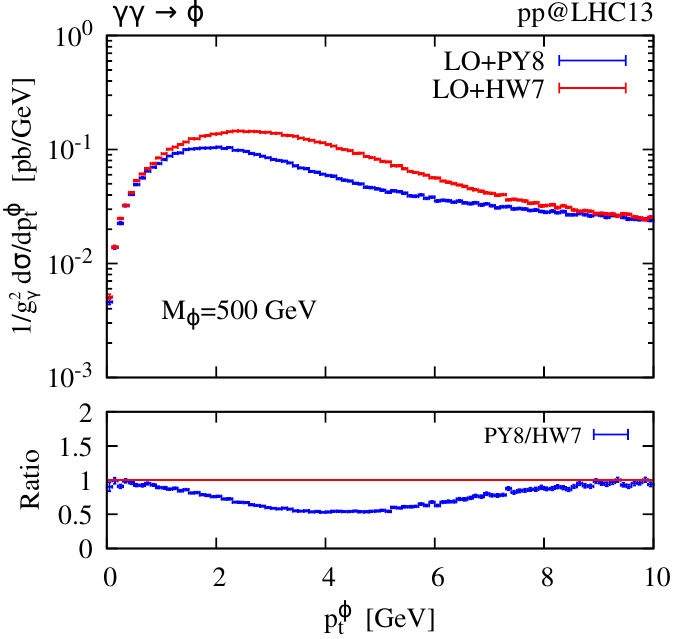}}
  \caption{\label{fig:yyh-PY8-HW7-pth} Transverse momentum spectrum of
    the scalar particle in proton-proton collisions at
    $\sqrt{s}=13\,$TeV for $M_\phi=500\,$GeV computed at leading
    order. The right panel shows the distributions at low $p_T$
    values.  The blue (red) line shows the prediction obtained with
    \Pythia{} (\Herwig{}).  }
\end{figure}

\begin{figure}[htb]
  \resizebox{7cm}{!}{\includegraphics{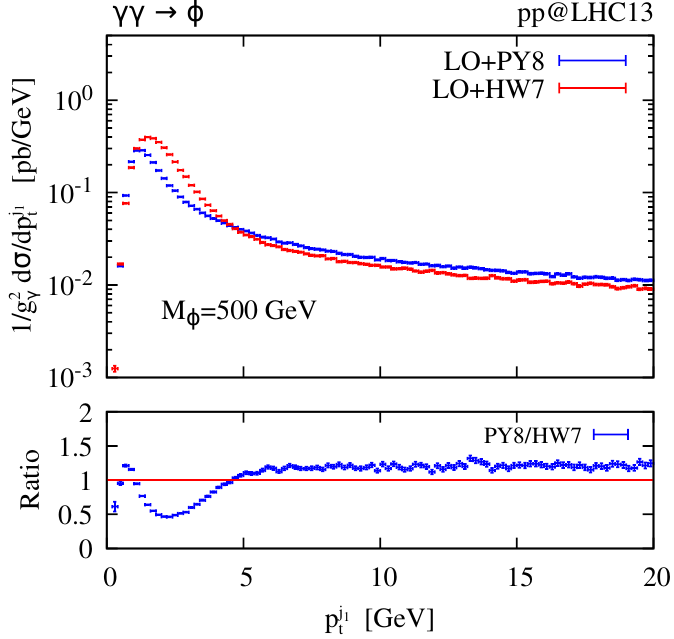}}
  \resizebox{7cm}{!}{\includegraphics{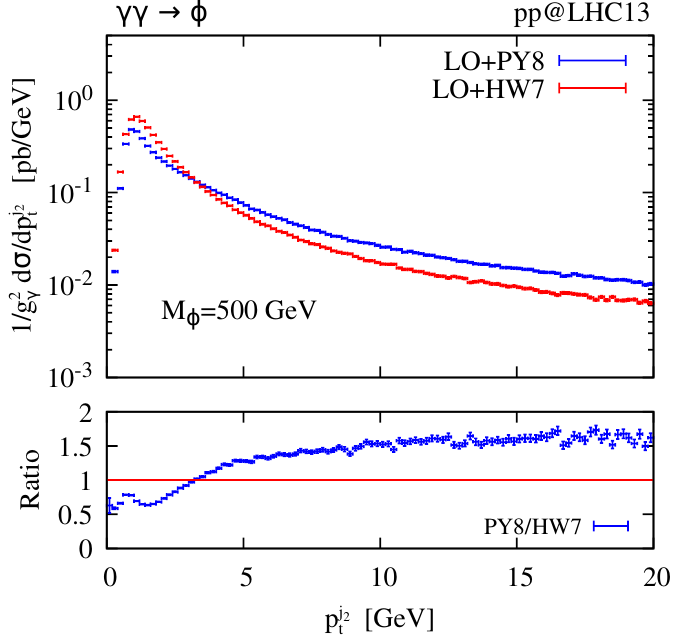}}
  \caption{\label{fig:yyh-PY8-HW7-ptjets} Transverse momentum spectrum
    of the leading (left panel) and second (right panel) jet in
    proton-proton collisions at $\sqrt{s}=13\,$TeV for
    $M_\phi=500\,$GeV computed at leading order.  The blue (red) line
    shows the prediction obtained with \Pythia{} (\Herwig{}).  }
\end{figure}

\begin{figure}[htb]
  \resizebox{7cm}{!}{\includegraphics{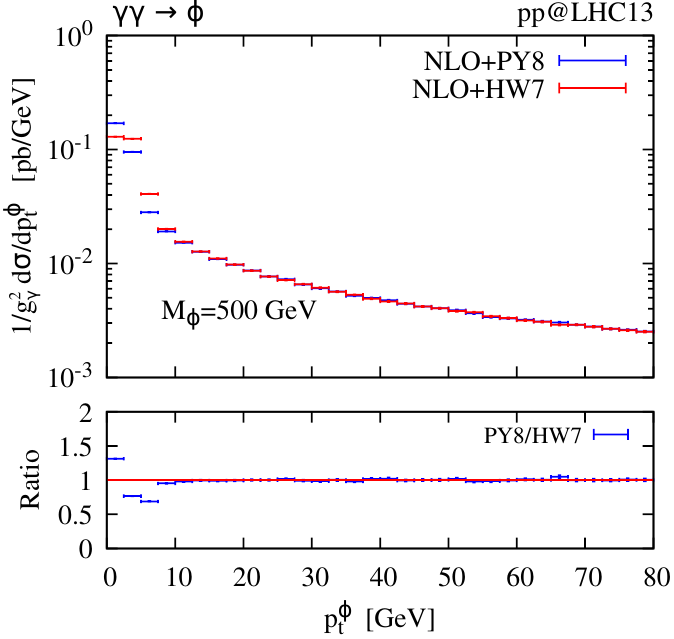}}
  \resizebox{7cm}{!}{\includegraphics{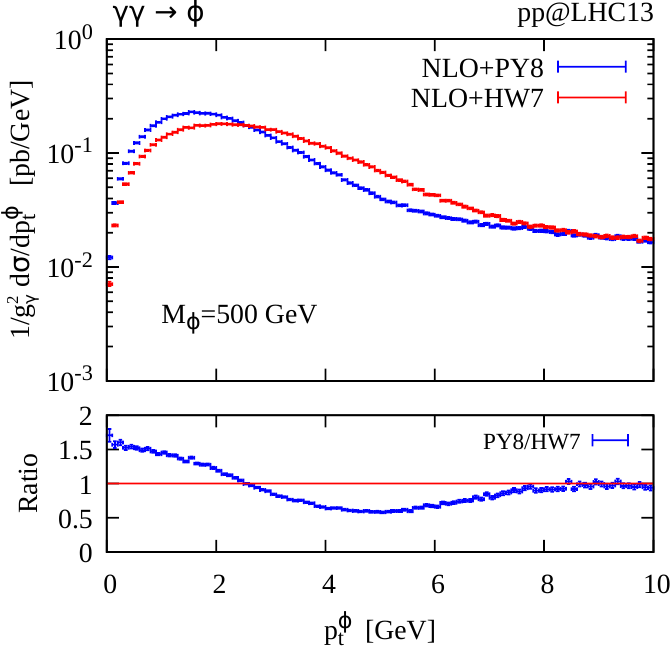}}
  \caption{\label{fig:yyh-PWG+PY8-PWG+HW7-pth}
    As in Fig.~\ref{fig:yyh-PY8-HW7-pth} computed at the NLO+PS level, comparing 
    the prediction obtained with \POWHEG+\Pythia{} (blue) with \POWHEG+\Herwig{} (red).
  }
\end{figure}

\begin{figure}[htb]
  \resizebox{7cm}{!}{\includegraphics{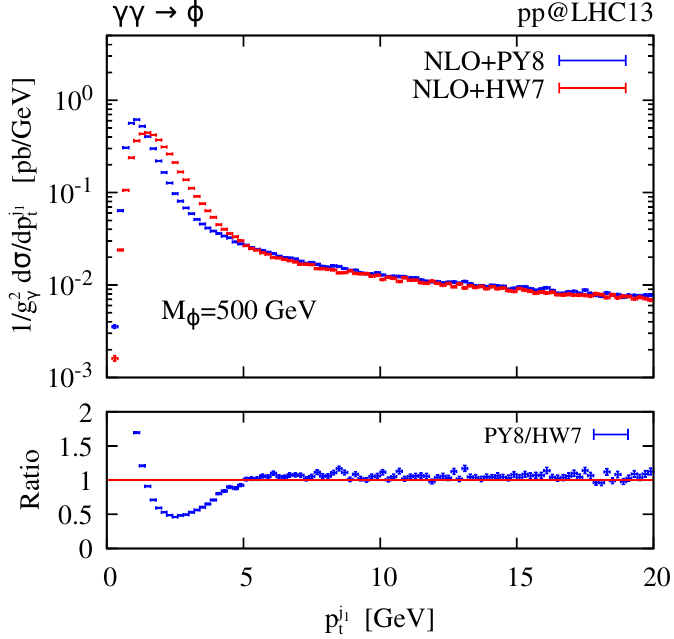}}
  \resizebox{7cm}{!}{\includegraphics{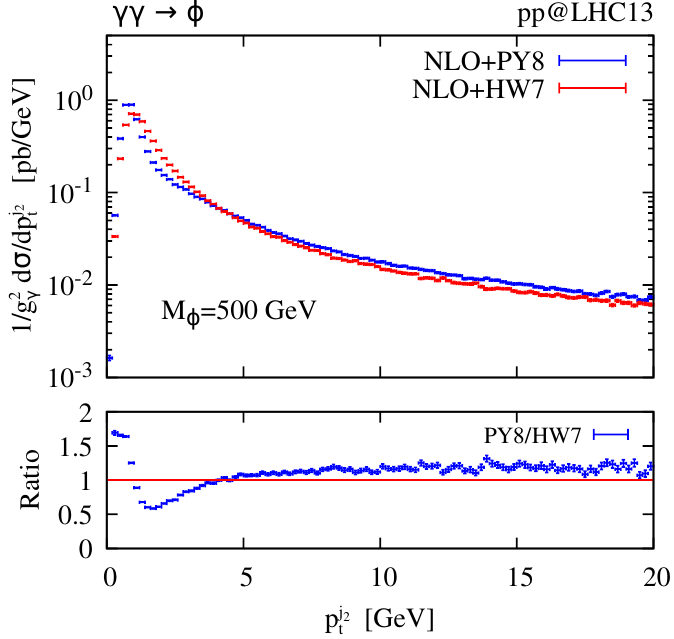}}
  \caption{\label{fig:yyh-PWG+PY8-PWG+HW7-ptjets} As in
    Fig.~\ref{fig:yyh-PWG+PY8-PWG+HW7-pth}, for the leading (left) and
    second (right) jet.  }
\end{figure}

In the following, we fix the mass of the scalar to $M_\phi = 500\,$GeV
and adopt the optimal scale choice $\mu_0=M_\phi$. We show results for
the transverse momentum of the resonance and of the leading and second
jets.  In Figs.~\ref{fig:yyh-PY8-HW7-pth}
and~\ref{fig:yyh-PY8-HW7-ptjets} we compare leading order \Pythia{}
and \Herwig{} predictions.
Notice that at this order, the selected observables are entirely
generated by the shower. We observe that \Pythia{} tends to produce
harder spectra with their peak shifted to lower $p_T$-values with
respect to \Herwig{}. Overall, the differences reach about $50\%$ in
both the low~$\pt$ and high $\pt$ regions. In the former, this can be
attributed to the differences among the shower and hadronization
approaches, and in the latter to the different modeling of hard
radiation away from the collinear region in the two generators.  The
same results including NLO corrections are displayed in
Figs.~\ref{fig:yyh-PWG+PY8-PWG+HW7-pth}
and~\ref{fig:yyh-PWG+PY8-PWG+HW7-ptjets}. We observe a better
agreement in the tails of the $p_T$ distributions of the resonance and
of the leading jet.  This is expected since these variables are now
modeled by the real corrections in the fixed-order
computation. Differences persist in the tail of the spectrum of the
second jet, although they are milder. The shapes at small $p_T$ values
are qualitatively similar to the pure shower results of
Figs.~\ref{fig:yyh-PY8-HW7-pth} and~\ref{fig:yyh-PY8-HW7-ptjets}.

\begin{figure}[htb]
  \resizebox{7cm}{!}{\includegraphics{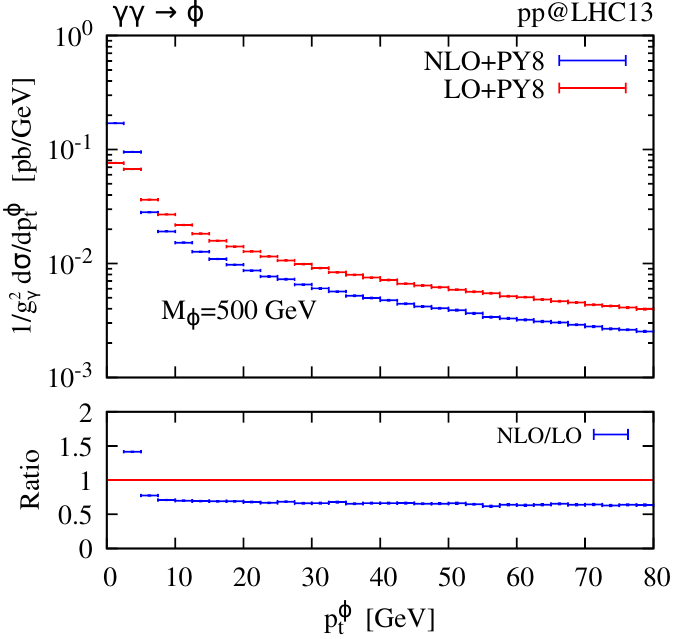}}
  \resizebox{7cm}{!}{\includegraphics{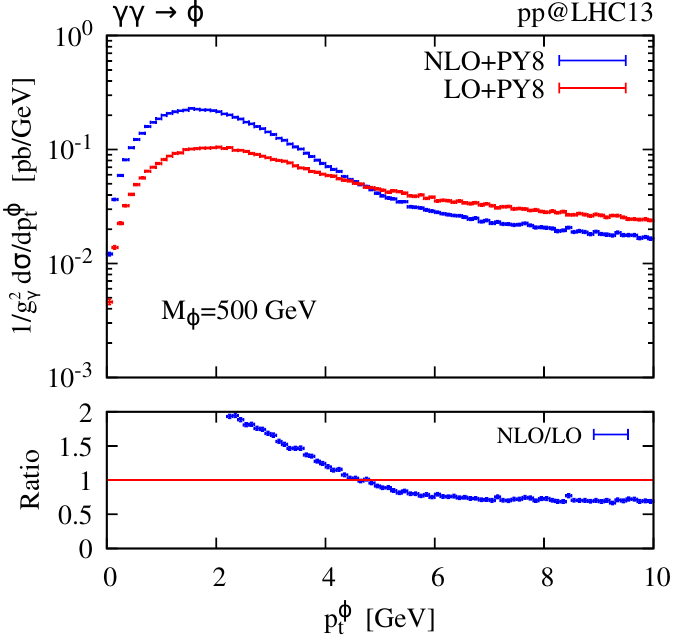}}
  \caption{\label{fig:yyh-PWG+PY8-PY8-pth} As in
    Fig.~\ref{fig:yyh-PY8-HW7-pth} comparing \POWHEG+\Pythia{} (blue)
    with \Pythia{} (red).  }
\end{figure}

\begin{figure}[htb]
  \resizebox{7cm}{!}{\includegraphics{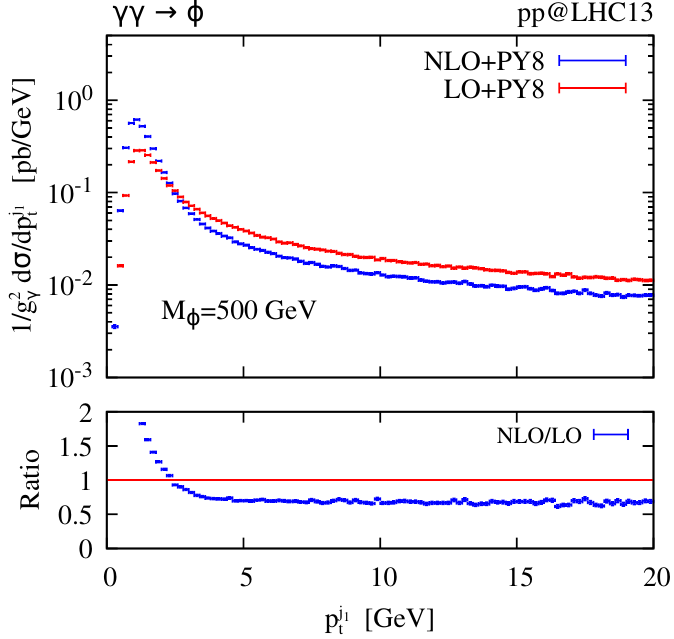}}
  \resizebox{7cm}{!}{\includegraphics{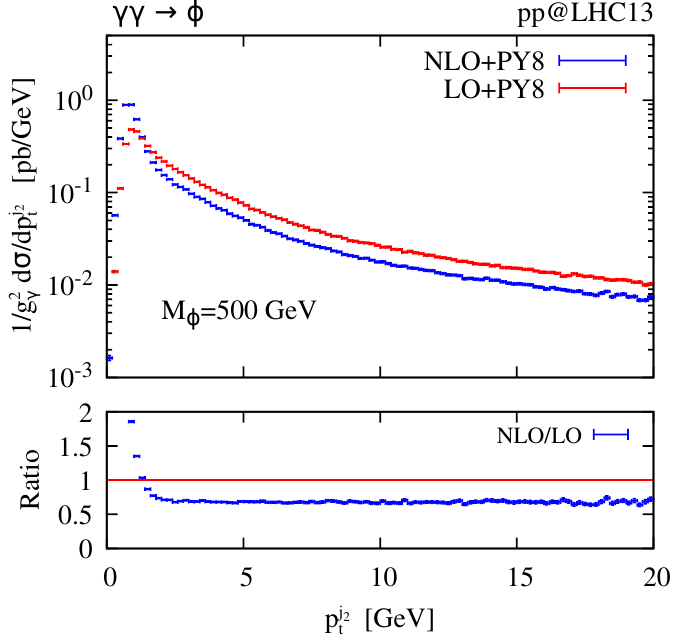}}
  \caption{\label{fig:yyh-PWG+PY8-PY8-ptjets} As in
    Fig.~\ref{fig:yyh-PY8-HW7-ptjets}, comparing \POWHEG+\Pythia{}
    (blue) with \Pythia{} (red).  }
\end{figure}

In Fig.~\ref{fig:yyh-PWG+PY8-PY8-pth} and
\ref{fig:yyh-PWG+PY8-PY8-ptjets},
we show a comparison between NLO+PS and LO+PS results using
\Pythia{}. We have obtained similar results also using \Herwig{}. We
see that the shower results provide a good qualitative description of
all the considered spectra, with rather flat K-factors for moderate to
high $p_T$-values. They tend to give harder distributions,
overshooting the NLO results by a factor of almost two for transverse
momenta larger than about $5\,$GeV.

Summarizing, we find considerable differences in the transverse
momentum distribution of the scalar when comparing \Pythia{} and
\Herwig{} at leading order, the \Pythia{} spectrum being considerably
harder. This feature is mitigated when NLO corrections are introduced,
and in fact very good agreement is found both in the large transverse
momentum distribution of the scalar and of the two hardest jets. On
the other hand, at very low transverse momenta (i.e. $p_T\lesssim
10$~GeV) the two shower generators yield a very different modeling of
the shape of the distributions. These features are not very relevant
when searching for heavy resonances, however they give an indication
that a more accurate tuning of the shower generators is needed for the
modeling of \elgaminit{} processes. We will return to this point after
the discussion of the lepton-scattering processes.

\subsection{Differential distributions for $e^+ e^- \to \phi$}
\label{sec:scalarfromleptons}
\begin{figure}[htb]
  \resizebox{7cm}{!}{\includegraphics{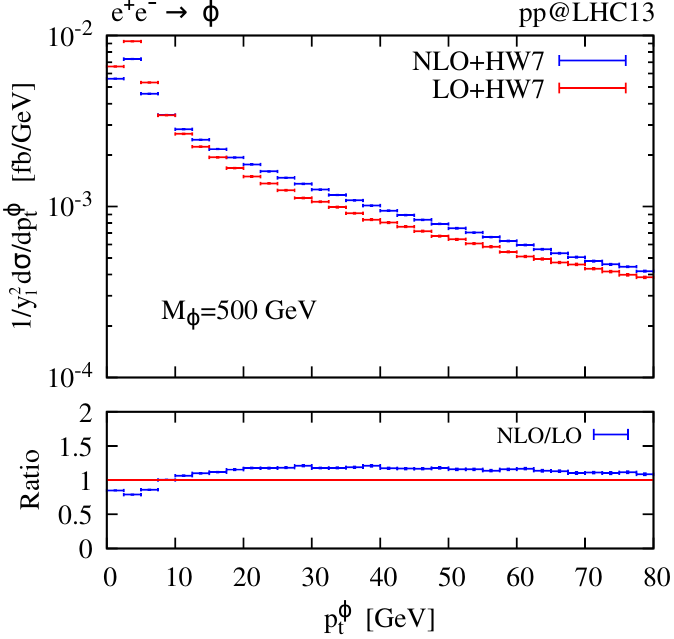}}
  \resizebox{7cm}{!}{\includegraphics{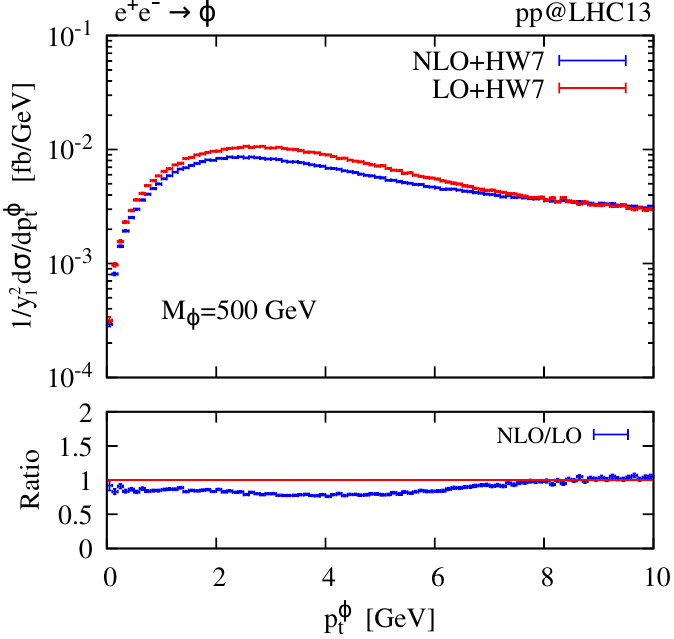}}
  \caption{\label{fig:llh-PWG+HW7-HW7-pth} Transverse momentum
    spectrum of the scalar particle in the \elinit{} production
    process at the $13\,$TeV LHC for $M_\phi=500\,$GeV. The right
    panel shows the distributions at low $p_T$ values.  We compare the
    pure \Herwig{} parton shower prediction (red) with the one
    obtained after the matching with the NLO computation (blue).  }
\end{figure}

\begin{figure}[htb]
  \resizebox{7cm}{!}{\includegraphics{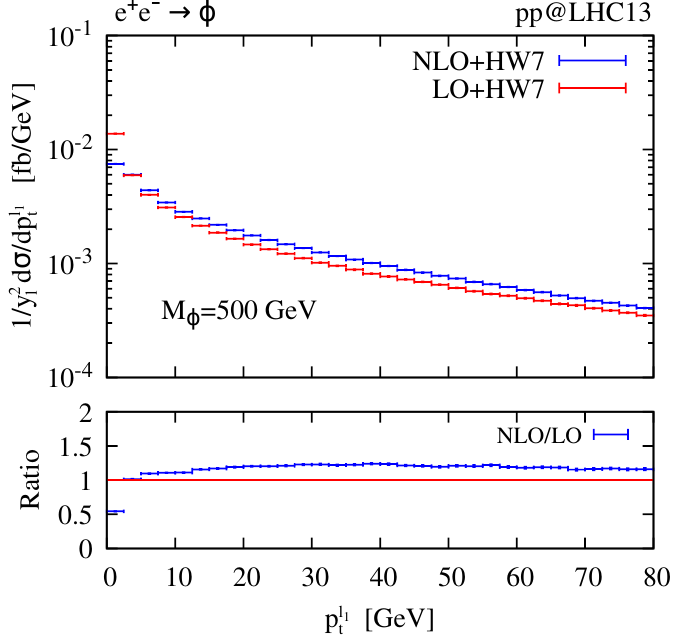}}
  \resizebox{7cm}{!}{\includegraphics{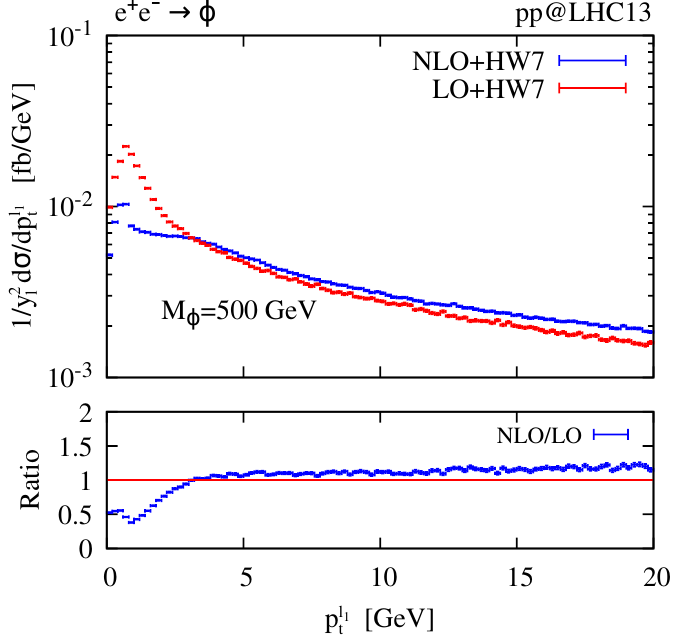}}
  \caption{\label{fig:llh-PWG+HW7-HW7-ptl1} Transverse momentum
    spectrum of the leading lepton in the \elinit{} scalar production
    process at the $13\,$TeV LHC for $M_\phi=500\,$GeV. The right
    panel shows the distributions at low $p_T$ values.  We compare the
    pure \Herwig{} parton shower prediction (red) with the one
    obtained after the matching with the NLO computation
    (blue).  }
\end{figure}

For this process we can only use \Herwig{} since \Pythia{} does not
currently support initial-state leptons.  In
Fig.~\ref{fig:llh-PWG+HW7-HW7-pth} and \ref{fig:llh-PWG+HW7-HW7-ptl1},
we show the transverse momentum of the scalar resonance and of the
leading lepton, comparing the pure shower results with the NLO+PS ones
using \Herwig{}.  In this case, we find a qualitatively good agreement
among them. At variance with the photon initiated process, the NLO
corrections lead to a harder spectrum in both distributions. Above
$20\,$GeV, we observe a rather flat K-factor of about 1.15.
In the region of small values of the transverse momentum of the
leading lepton, we observe a discontinuity in the NLO+PS prediction
related to the minimum $p_T$ allowed in the \POWHEG{} generation.
Again, this is not a crucial problem in the simulation of the
production of high-mass states.  Nonetheless, it shows the need for
further investigation in the modeling of \elgaminit{} processes by
the showers.

\begin{figure}[htb]
  \resizebox{7cm}{!}{\includegraphics{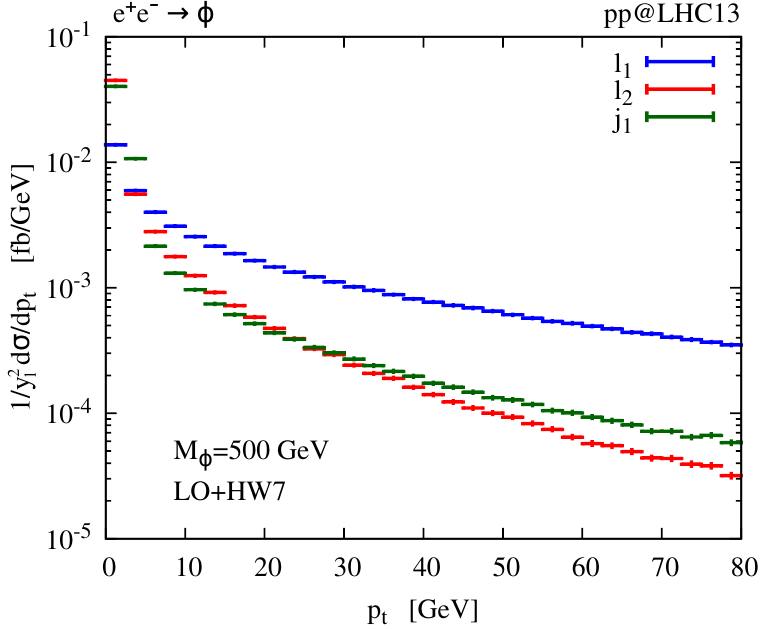}}
  \resizebox{7cm}{!}{\includegraphics{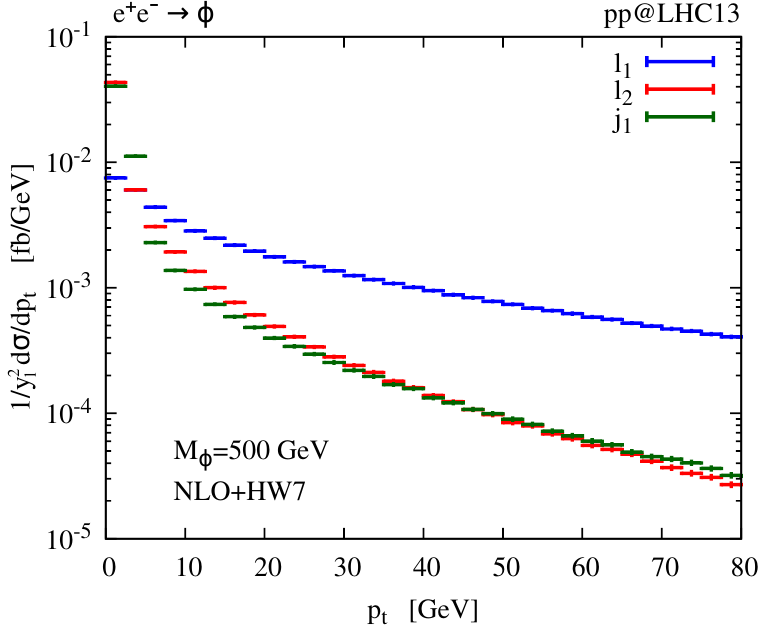}}
  \caption{\label{fig:llh-pt-l1-l2-j1} Transverse momentum spectrum of
    the leading lepton, second lepton and leading jet in the \elinit{}
    scalar production process at the $13\,$TeV LHC for
    $M_\phi=500\,$GeV as obtained with \Herwig{} (left panel) and
    after matching at NLO (right panel).}
\end{figure}

The shower allows us to investigate the associated hadronic
activity in the event.  In Fig.~\ref{fig:llh-pt-l1-l2-j1}, we compare
the transverse momentum of the leading leptons (hardest and
next-to-hardest) with that of the leading jet. Notice that at LO+PS
all the emissions are provided by the shower, while at NLO+PS the
hardest emission (leading lepton) is generated according to the exact
matrix element. The second lepton and leading jet, being
next-to-hardest emissions, display a much softer spectrum with respect
to the leading lepton, and are comparable.  The shower results present
a slightly harder $p_t$ jet spectrum, while the differences are much
milder in the NLO+PS computation.  The overall picture of the
next-to-hardest radiation is consistent with backward evolution
starting from a photon and a lepton leg. The probability to have a
$q\to q \gamma$ splitting for the first and a $\gamma\to l {\bar l}$
for the second are in fact parametrically of the same order.

\section{Lepton scattering}
\label{sec:lepscat}
We are interested in lepton scattering processes such that the
dilepton signal does not have excessive competition from other more
abundant Standard Model (SM) processes. As a minimum requirement, we
must focus upon final states not affected by the very large Drell-Yan
background. Thus, we consider lepton pairs of different flavours or
with the same electric charge. In the following we will refer to these
signatures as non-Drell-Yan (NDY) pairs.  We have implemented the NDY
processes in the \POWHEGBOX{} framework, in such a way that, besides
being able to compute the NLO cross sections we can also generate
events to be interfaced with a shower Monte Carlo program.

As we will see in the following, the NDY processes may be observable
at the LHC as long as the transverse momenta of the leptons are not
too large. Thus both higher-order effects and shower details can
significantly affect the result. Moreover, lepton-initiated processes
have become available in shower programs only very recently, and their
implementation is still in a preliminary stage. For these reasons we
have also implemented a direct calculation of four lepton production
using the \HT{} (Hadronic Tensor) method, in order to better assess
the reliability of our approach, and to provide a guide for the tuning
of the Monte Carlos. The \HT{} calculation is carried out by first
computing the matrix element for the production of two lepton pairs,
initiated by two virtual photons. This matrix element is contracted
with the hadronic tensors of each incoming proton, and a full phase
space integration is performed. This calculation is described in
detail in appendix~\ref{app:HT}.

We stress that in our \HT{} calculation we include all contributions
to the photon-induced four-leptons production, at order ${\cal
  O}(\alpha^4)$.  This is in contrast with the calculation of
Ref.~\cite{Harland-Lang:2021zvr}, where diagrams of this order with
the two incoming photons attached to the same fermion line are instead
neglected. These contributions, not included in
Ref.~\cite{Harland-Lang:2021zvr}, enter already in our NLO calculation
(see Fig.~\ref{fig:singregs}, (c) and (d)) and as
discussed later, are not completely negligible. Furthermore,
interference terms in the case of identical fermions are never
included in Ref.~\cite{Harland-Lang:2021zvr}.

The \HT{} calculation includes all higher-order QCD corrections to the
NDY processes. On the other hand it does not include the effect of
initial-state collinear radiation of photons from leptons. These
effects are of the order of the electromagnetic coupling constant
$\aem$ multiplied by a collinear logarithm $L=\log{\pt}{/\Lambda}$,
i.e. of order $\aem/\as$, that in our counting (that is $\aem \approx
\as^2$) should be considered equivalent to an NLO correction. Rather
than trying to add these effects to the \HT{} calculation, we have
estimated their size by switching them off in the calculation using
the lepton PDFs by suitably adapting the \hoppet{}~\cite{Salam:2008qg}
routines used in the \LUXlep{} paper. We have found that at the
transverse momenta that we have considered these effects are below the
percent level, and thus we can trust the \HT{} result for a validation
of our calculations.\footnote{However, for the production of high-mass
  objects, these effects become larger. We checked that for scalar
  production by lepton collisions they amount to a reduction of
  1\%{} (2.6\%{}) of the cross section for a scalar with mass equal to
  125 (1000) GeV.}  We stress however that the \HT{} calculation is in
general much more demanding than the \POWHEG{} NLO one. Furthermore,
it cannot be straightforwardly interfaced to a shower program, and thus we have
not attempted to turn it into a full event generator.\footnote{For
  an approach to build such an interface see~\cite{Harland-Lang:2020veo}.}

\subsection{The NLO POWHEG calculation}
\label{sec:NLOPWGcalc}
\begin{figure}[h]
  \centering \includegraphics[width=0.28\textwidth]{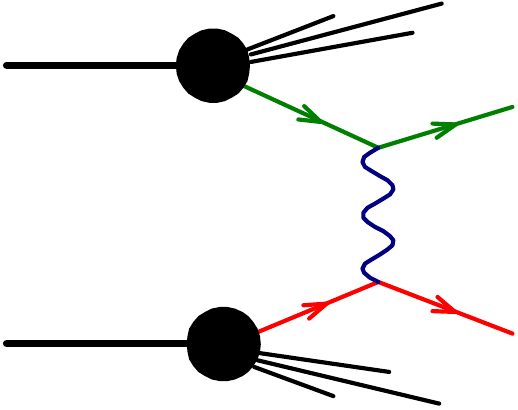}
  \hskip 1.5cm
   \includegraphics[width=0.32\textwidth]{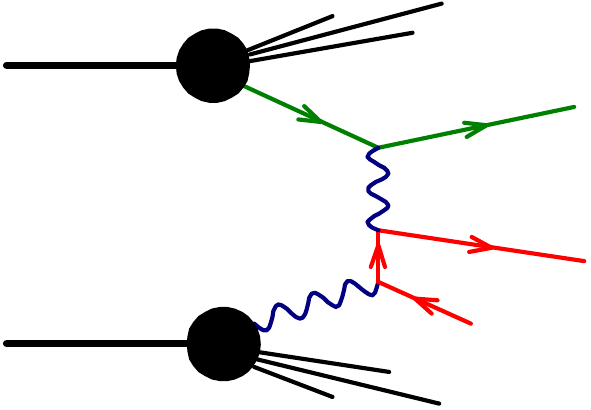}
   \caption{\label{fig:BornReal}
     Left: the Born level graph for same-sign, different-flavours
     lepton scattering; right: an example of a real graph associated
    with radiative corrections to this Born graph.}
\end{figure}

\begin{figure}[htb]
  \centering
  \begin{tabular}{cccc}
  \includegraphics[width=3cm]{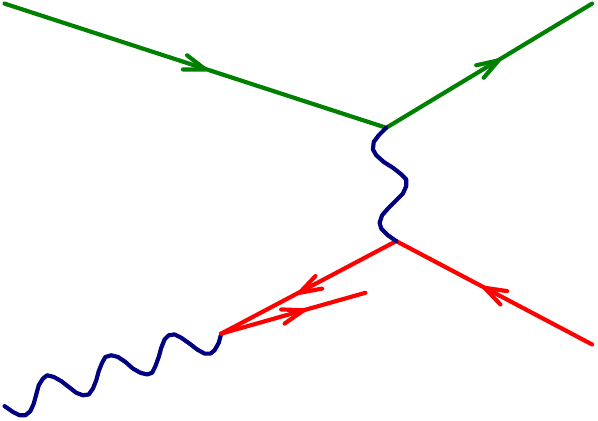} &
  \includegraphics[width=3cm]{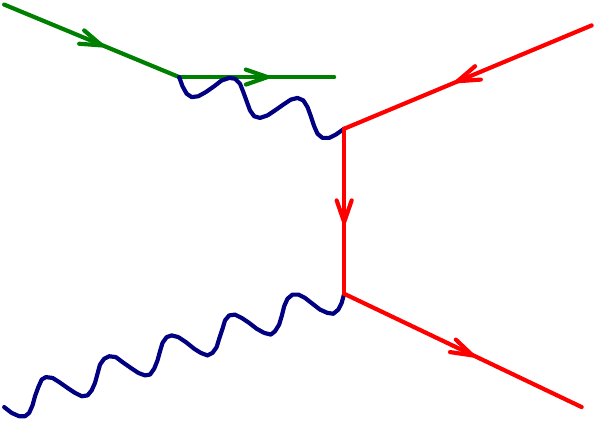} &
  \includegraphics[width=3cm]{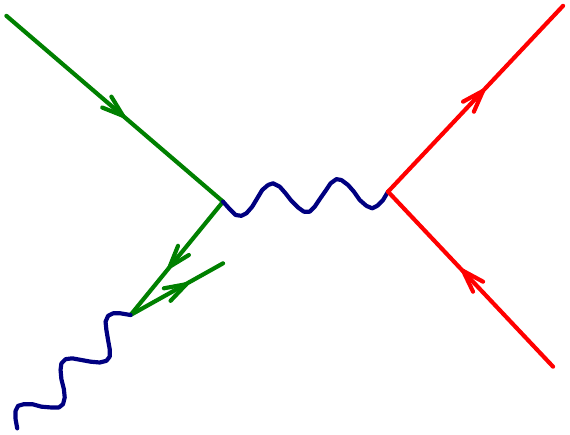} &
  \includegraphics[width=3cm]{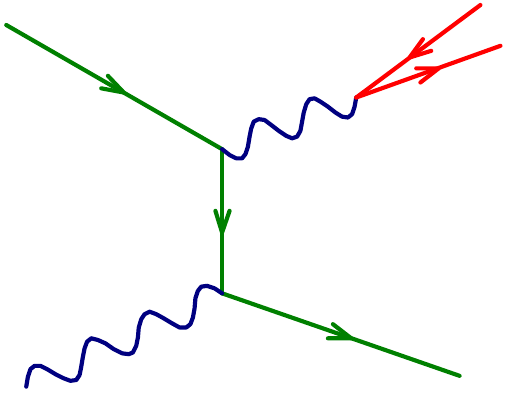} \\
    (a) & (b) & (c) & (d)
  \end{tabular}
  \caption{\label{fig:singregs} Singular regions associated with real
    graphs in the lepton scattering process.}
\end{figure}

In the left part of Fig.~\ref{fig:BornReal}, we display
one of the Feynman graph for the process of same-sign, different-flavours
scattering at the Born level.
We also consider same-sign identical-flavours and opposite-sign
different-flavours.  In the case of identical flavours an exchange
graph must also be included.  The NLO corrections to these processes
arise from the graphs where one incoming lepton originates from an
incoming photon, as shown in the right part of Fig.~\ref{fig:BornReal}.
The real graphs for the NLO corrections have several singular regions,
that are shown schematically in Fig.~\ref{fig:singregs},
where, for the sake of definiteness, we focus upon the scattering of
leptons with different flavour.  The singular regions are
\begin{enumerate}[(a)]
\item A final-state fermion with flavour different from the incoming
  one becomes collinear to the incoming photon.  The underlying Born
  for this singular region corresponds to a process of
  different-flavours lepton scattering, i.e. the same process that we
  want to include at the Born level.  The corresponding collinear
  singularities are removed following the usual factorization
  procedure.
\item A final-state fermion of the same flavour of the initial one
  becomes collinear to it.  This singular region yields a final state
  that is Drell-Yan like. However, this graph does contribute at the
  NLO level to our signal when far away from the singular region, so
  it must be included. Its underlying Born configuration is $\gamma \gamma
  \rightarrow l^+ l^-$.
\item A final-state fermion with flavour equal to the incoming one
  becomes collinear to the incoming photon. Also in this case, the
  underlying Born configuration is Drell-Yan like, but away from the singular region
  it may contribute at the NLO level to our signal, and must be
  included.
\item Two opposite-flavour final-state leptons become collinear to
  each other. This process can generate a different-flavour pair that
  is not Drell-Yan like. Yet, it cannot be ascribed to a lepton
  scattering process, since its underlying Born is the Compton process
  $\gamma l \to \gamma l$.  We must however include it since it is
  relevant at the NLO level.  In order to suppress this contribution
  we will impose an isolation criterion upon our final leptons.
\end{enumerate}
Configuration (a) is dealt with automatically by the \POWHEGBOX{}
framework.  In order to deal with the remaining singular regions, we
introduce also the Born subprocesses $\gamma \gamma \rightarrow l^+
l^-$, $l^+ l^- \rightarrow \ell^{_{} +} \ell^-$ and $l \gamma
\rightarrow l \gamma$. In this way, regions (b), (c), and (d) are
correctly detected as singular ones. We then use the Bornzerodamp feature of the
\POWHEGBOX~\cite{Melia:2011gk} to separate those regions
as remnants and treat them as pure unsubtracted\footnote{This is obtained setting to zero the
corresponding Born matrix elements.} real contributions.
We notice that the divergent contributions of regions (b) and (c) are
eventually excluded by the analysis cuts. On the other hand, a realistic cut to
exclude region (d) would require complete isolation of the
leptons. Since lepton detection cannot go below transverse momenta of
the order of a GeV, a complete isolation cannot be achieved.
Thus, some logarithmic sensitivity to the lepton mass remains, and
we must modify the matrix elements by including mass effects in diagrams that
involve the splitting of a photon into a lepton-antilepton pair. The
details of our procedure are given in appendix~\ref{app:sfsplitmass}.

In order to deal with the remnants corresponding to regions (b), (c) and (d),
that have a divergent or a logarithmically enhanced cross section,
we implement their production in \POWHEG{}
with a reduced sampling rate by applying a suitable suppression
factor, that is divided out from the weight of the generated event.
With this procedure, if the analysis cuts are applied, we obtain
correct, finite results.

\subsection{NLO results and scale dependence}
\label{sec:lepscatNLO}
We begin by studying the scale dependence of the NLO cross section for
all combinations of different-flavours or equal-sign lepton
scattering. For completeness, we include also the contribution of $Z$
exchange and $Z\gamma$ interference.  We notice that, at NLO,
scattering processes of leptons of different flavours and equal or
opposite charges have different cross sections even if only the
electromagnetic interaction is considered. In fact, the amplitude has
contributions of order $q_l q_\ell^2$ and $q_\ell q_l^2$, with $q_l$
and $q_\ell$ denoting the electric charges of the two fermions. These
two contributions change relative signs if one lepton is replaced with
its antilepton.  On the other hand, fully charge-conjugated processes
yield the same cross sections and distributions even if we include $Z$
exchange, provided we do not look at parity-sensitive
observables. Thus we can omit charge-conjugate processes.

In our analysis we always consider proton-proton collisions at 13~TeV.

As discussed in the previous section, at the NLO level we must impose
an isolation cut on the final-state leptons, in order to avoid the
collinear divergence of the region (d).  We thus require that no other
lepton with transverse momentum larger than a given cut is inside a
cone in the pseudorapidity-azimuthal plane with given aperture $\Delta
r$. We have considered the values $\Delta r=0.1,\,0.3$ and $0.6$. We
have observed a mild dependence on $\Delta r$, thus we present here
only the results for $\Delta r =0.3$.  Altogether, we require both
leptons to satisfy the cuts
\begin{equation}
  \begin{split}
  &\pt > p_{\rm cut},\; p_{\rm cut}=5, 10 \mbox{ and } 20\,  {\rm GeV}\,, \qquad |\eta| < 2.4\,,\\
  &\mbox{isolation:}\quad \Delta r>0.3\quad \mbox{for leptons}\;  \pt > 0.9\,{\rm GeV}\,.
  \end{split}
  \label{eq:cuts}
\end{equation}

\begin{figure}[htb]
  \centering
  \includegraphics[width=14cm]{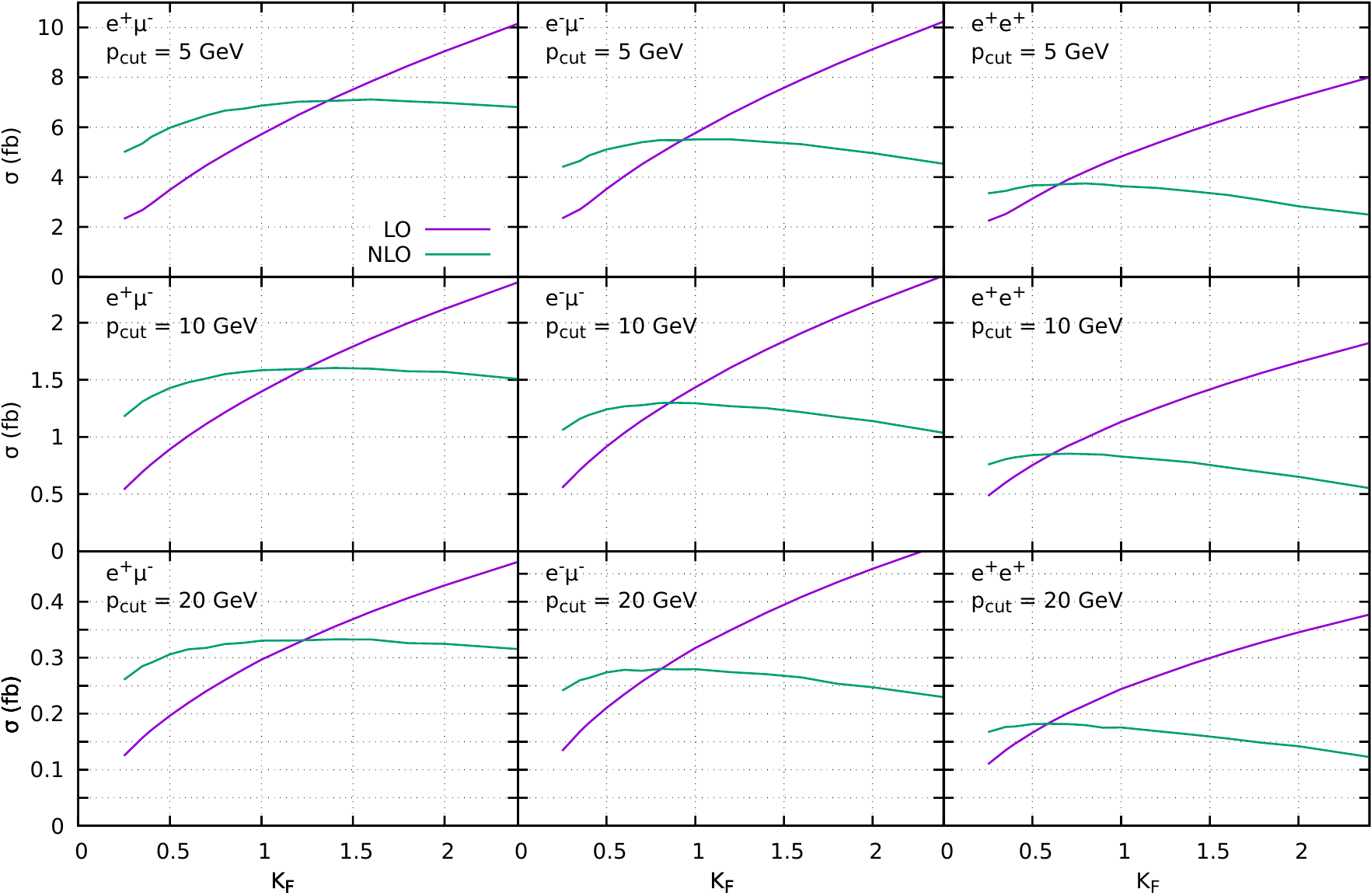}
  \caption{\label{fig:scalescan1} Scale dependence of the total cross
    section for lepton scattering at LO and NLO for $e^+\mu^-$,
    $e^-\mu^-$ and $e^+e^+$ production for different $p_t$ cuts.}
\end{figure}

\begin{figure}[htb]
  \centering
  \includegraphics[width=14cm]{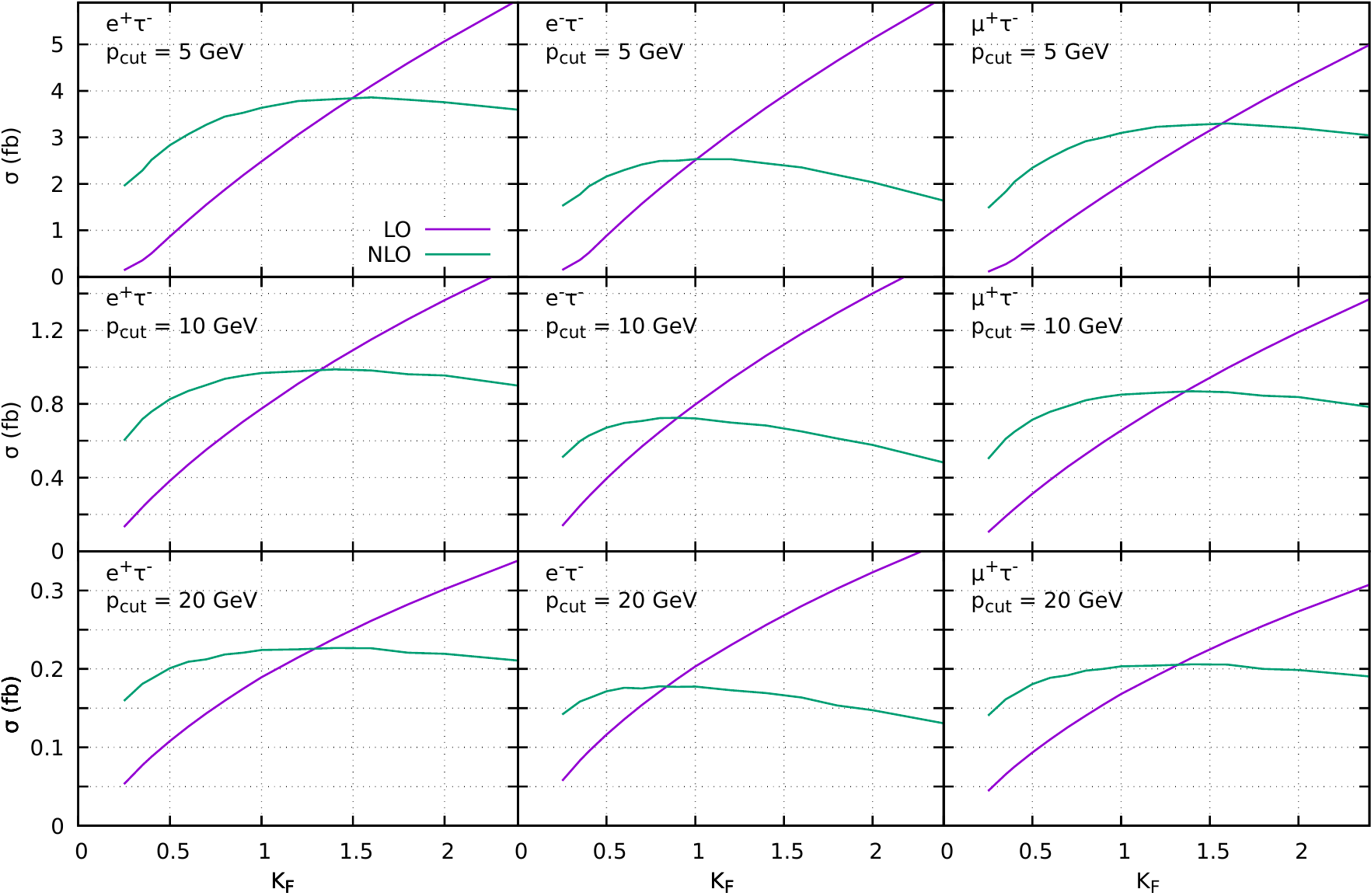}
  \caption{\label{fig:scalescan2} As Fig.~\ref{fig:scalescan1} for
    $e^+\tau^-$, $e^-\tau^-$ and $\mu^+\tau^-$.}
\end{figure}

\begin{figure}[htb]
  \centering
  \includegraphics[width=14cm]{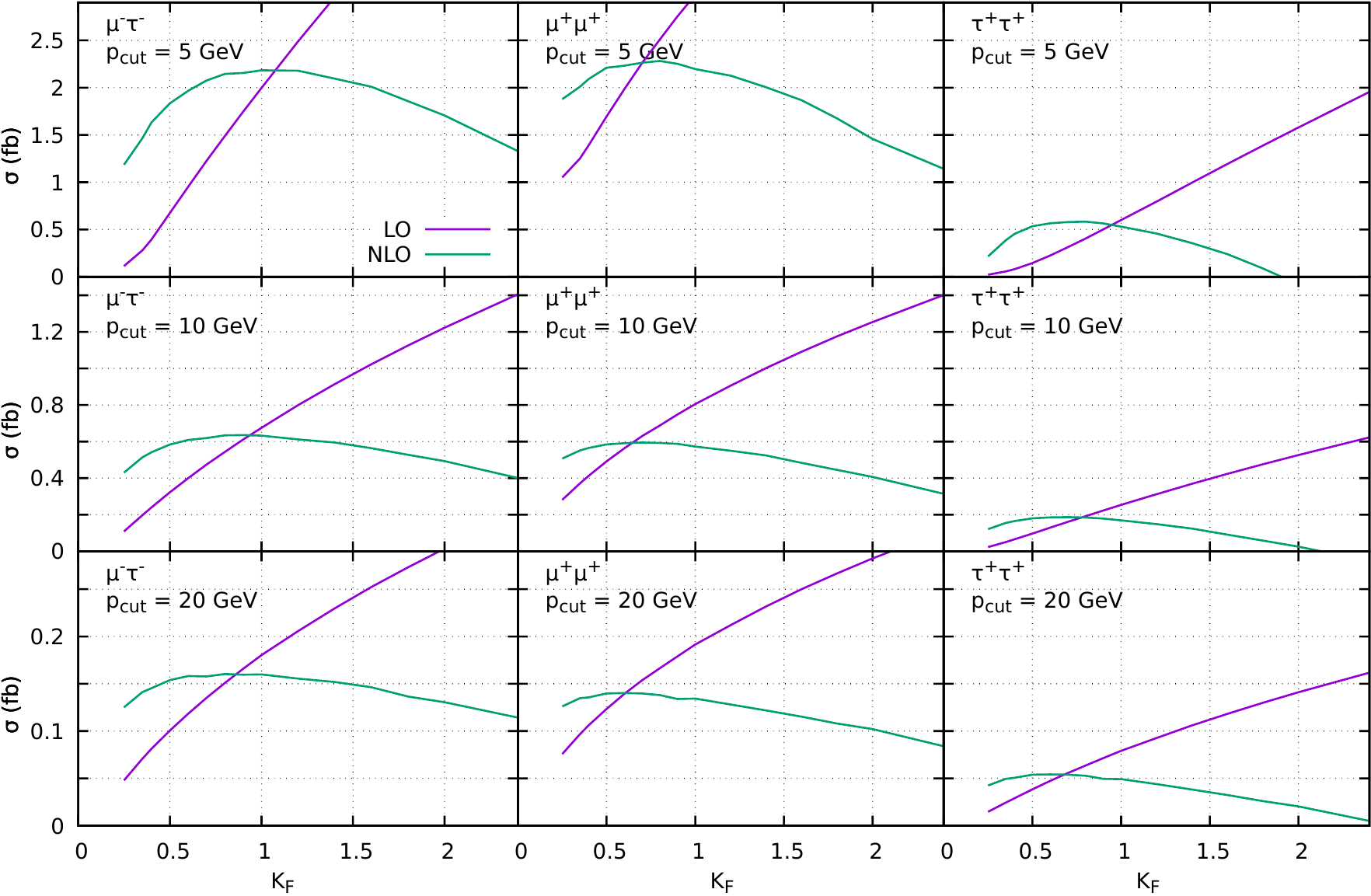}
  \caption{\label{fig:scalescan3} As Fig.~\ref{fig:scalescan1} for
    $\mu^-\tau^-$, $\mu^+\mu^+$ and $\tau^+\tau^+$.}
\end{figure}

In figures~\ref{fig:scalescan1}, \ref{fig:scalescan2} and
\ref{fig:scalescan3}
we show the cross section as a function of the factorization scale factor
$K_F=\mu_F/\mu_0$ for different choices of the transverse momentum cut.
We have chosen the reference scale $\mu_0$ as
the sum of the absolute value of the transverse momentum of all
final-state leptons divided by two. This choice is collinear-safe,
since for both initial- and final-state collinear singularities it
reduces to the transverse momentum of the final-state particles in the
underlying Born process. We notice that, in general, there is a
considerable reduction in the scale dependence when going from the LO
to the NLO result. Furthermore, we observe that generally the
intersection of the LO and NLO results is slightly above 1 for
opposite-sign, different-flavour leptons, slightly below 1 for
same-sign, different-flavour, and near 0.5 for same-sign same-flavour
leptons.  While the scale behaviour of the results for the electrons
and muons seems very reasonable, when the $\tau$ is involved, we see
an anomalously large scale dependence even at the NLO level, the worse
case being $\tau\tau$ scattering. This behaviour is partly understood
as a consequence of the fact that for small scales the $\tau$ parton
density becomes negative, and in fact we see that for small scales the
LO cross section becomes very small.

Given the differences in the scale behaviour of the different
processes, we have chosen the following procedure for selecting a
range of scales appropriate to each process. We have scanned the cross
sections for $0.5<K_F<4$, with a spacing of 0.1 for $0.5<K_F<1$, 0.2
for $1<K_F<2$, and 0.4 for $K_F>2$.  We have determined the value of
$K_F$ where the NLO cross section is maximal, and taken half and twice
this $K_F$ value as extremes for the scale range. We then quote the
maximum and minimum value of the cross section in this range, computed
at both the LO and NLO level.  The results are shown in
table~\ref{tab:dileptonNLOsigmas}, where we also report the
predictions obtained with the \HT{} method.
\begin{table}[htb]
  \centering
\begin{tabular}{|c|c|cc|cc|cc|c|}
\hline
 Proc. & $p_{\rm cut}$ & \multicolumn{2}{|c|}{$K_{\rm F}$} &\multicolumn{2}{|c|}{NLO} & \multicolumn{2}{|c|}{LO} & \HT{} \\
\hline
& & min & max  & min & max & min & max & \\
\hline
\hline
           & $5$ &   $ 0.80$ & $ 3.20$ & $ 6.203$ & $ 7.113$ & $ 4.907$ & $12.091$ & $7.41$      \\  
\cline{2-9}
$e^+\mu^-$ & $10$ &   $ 0.70$ & $ 2.80$ & $ 1.451$ & $ 1.604$ & $ 1.116$ & $ 2.563$ &  $1.67$     \\  
\cline{2-9}
          & $20$ &   $ 0.70$ & $ 2.80$ & $ 0.305$ & $ 0.333$ & $ 0.241$ & $ 0.510$ &  $0.346$     \\  
\hline
\hline
       & $5$ &   $ 0.50$ & $ 2.00$ & $ 4.961$ & $ 5.515$ & $ 3.521$ & $ 9.127$ &  $6.65$     \\  
\cline{2-9}
$e^-\mu^-$       & $10$ &   $ 0.40$ & $ 1.80$ & $ 1.176$ & $ 1.299$ & $ 0.784$ & $ 2.046$ &  $1.53$     \\  
\cline{2-9}
       & $20$ &   $ 0.40$ & $ 1.60$ & $ 0.264$ & $ 0.280$ & $ 0.183$ & $ 0.409$ &  $0.322$     \\  
\hline
\hline
       & $5$ &   $ 0.80$ & $ 3.20$ & $ 3.048$ & $ 3.860$ & $ 1.875$ & $ 7.549$ &  3.48     \\  
\cline{2-9}
 $e^+\tau^-$      & $10$ &   $ 0.70$ & $ 2.80$ & $ 0.846$ & $ 0.988$ & $ 0.553$ & $ 1.737$ &  0.962     \\  
\cline{2-9}
       & $20$ &   $ 0.70$ & $ 2.80$ & $ 0.200$ & $ 0.227$ & $ 0.143$ & $ 0.371$ &  0.228     \\  
\hline
\hline
       & $5$ &   $ 0.60$ & $ 2.40$ & $ 1.639$ & $ 2.531$ & $ 1.237$ & $ 6.011$ &  3.25     \\  
\cline{2-9}
$e^-\tau^-$       & $10$ &   $ 0.40$ & $ 1.80$ & $ 0.613$ & $ 0.725$ & $ 0.298$ & $ 1.295$ &  0.871     \\  
\cline{2-9}
       & $20$ &   $ 0.40$ & $ 1.60$ & $ 0.163$ & $ 0.178$ & $ 0.095$ & $ 0.280$ &  0.215     \\  
\hline
\hline
       & $5$ &   $ 0.80$ & $ 3.20$ & $ 2.518$ & $ 3.300$ & $ 1.469$ & $ 6.416$ & 2.74      \\  
\cline{2-9}
$\mu^+\tau^-$       & $10$ &   $ 0.70$ & $ 2.80$ & $ 0.732$ & $ 0.870$ & $ 0.459$ & $ 1.534$ & 0.816      \\  
\cline{2-9}
       & $20$ &   $ 0.70$ & $ 2.80$ & $ 0.180$ & $ 0.206$ & $ 0.126$ & $ 0.339$ &  0.204     \\  
\hline
\hline
       & $5$ &   $ 0.50$ & $ 2.00$ & $ 1.706$ & $ 2.184$ & $ 0.674$ & $ 4.249$ &   2.53    \\  
\cline{2-9}
$\mu^-\tau^-$       & $10$ &   $ 0.40$ & $ 1.80$ & $ 0.527$ & $ 0.635$ & $ 0.240$ & $ 1.125$ &  0.742     \\  
\cline{2-9}
       & $20$ &   $ 0.40$ & $ 1.60$ & $ 0.145$ & $ 0.160$ & $ 0.081$ & $ 0.252$ &  0.187     \\  
\hline
\hline
       & $05$ &   $ 0.40$ & $ 1.60$ & $ 3.285$ & $ 3.745$ & $ 2.719$ & $ 6.338$ &  $4.9$     \\  
\cline{2-9}
$e^+e^+$       & $10$ &   $ 0.35$ & $ 1.40$ & $ 0.777$ & $ 0.853$ & $ 0.602$ & $ 1.363$ &  $1.05$     \\  
\cline{2-9}
       & $20$ &   $ 0.25$ & $ 1.20$ & $ 0.167$ & $ 0.182$ & $ 0.110$ & $ 0.267$ &  $0.213$     \\  
\hline
\hline
       & $5$ &   $ 0.40$ & $ 1.60$ & $ 1.868$ & $ 2.282$ & $ 1.394$ & $ 4.198$ &  2.82     \\  
\cline{2-9}
$\mu^+\mu^+$       & $10$ &   $ 0.35$ & $ 1.40$ & $ 0.524$ & $ 0.595$ & $ 0.371$ & $ 1.002$ & 0.711      \\  
\cline{2-9}
       & $20$ &   $ 0.25$ & $ 1.20$ & $ 0.126$ & $ 0.140$ & $ 0.076$ & $ 0.212$ &  0.163     \\  
\hline
\hline
       & $5$ &   $ 0.40$ & $ 1.60$ & $ 0.239$ & $ 0.583$ & $ 0.082$ & $ 1.193$ &  $0.610$     \\  
\cline{2-9}
$\tau^+\tau^+$       & $10$ &   $ 0.35$ & $ 1.40$ & $ 0.124$ & $ 0.187$ & $ 0.050$ & $ 0.369$ & $0.223$      \\  
\cline{2-9}
       & $20$ &   $ 0.25$ & $ 1.20$ & $ 0.043$ & $ 0.054$ & $ 0.015$ & $ 0.093$ &  $0.0658$     \\  
\hline
\end{tabular}
\caption{Cross sections in femtobarns for lepton scattering with
  acceptance cuts $|\eta|<2.4$ and $\pt> p_{\rm cut}$ for both
  leptons.  An isolation cut is also imposed by requiring no leptons
  in a cone of radius $r<0.3$ with $p_t> 0.9$ GeV.  For
  each subprocess, a range in the factorization scale factor $K_F$ is
  chosen by determining the value of $K_F$ that maximizes the NLO
  cross section and then taking half and twice this value as
  extremes, as detailed in the text.
  The quoted cross sections correspond to the minimum and
  maximum in this range.\label{tab:dileptonNLOsigmas}}
\end{table}
The range of the LO results are quite large, while it becomes more
modest in the NLO case. The NLO result is always included in the LO
band. We observe that the cross sections for like-sign different
fermions are substantially different than the unlike-sign ones. We
remark that the range of values given here for the LO results differs slightly
from the one given in the \LUXlep{} paper, because of the different
choice of scales.

We notice that, in comparison, the HT result is almost always larger
than the upper limit of the NLO band (apart for $e^+\tau^-$ and  $\mu^+\tau^-$), although not by large amounts.
We will see in the following, however, that the NLO+PS results are in
much better agreement with the \HT{} ones.  We remind the reader that
fixed-order NLO results in 2$\to$2 parton scattering are typically not
well-behaved also in the case of coloured partons, while \POWHEG{}
results perform much better~\cite{Alioli:2010xa}. Furthermore, we
remark that, in our case, the NLO+PS implementation recovers in an
approximate way effects that are included in the \HT{} result, but not
included at fixed order NLO.

\subsection{Study with realistic cuts}
\label{sec:cuts}
\begin{figure}[h]
  \centering
  \includegraphics[width=0.35\textwidth]{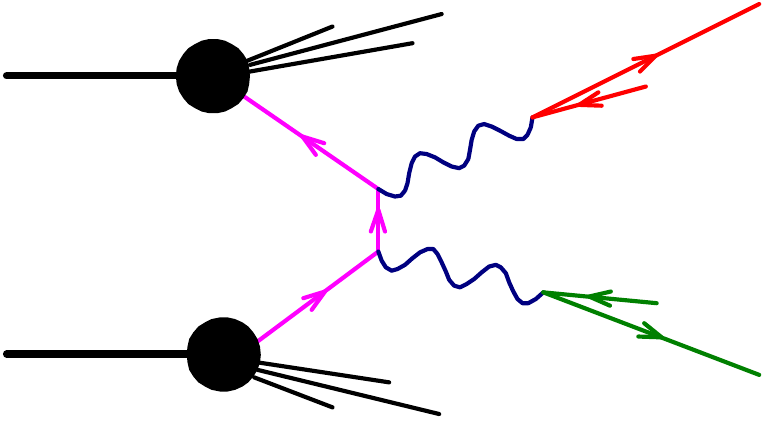}
  \caption{A double Drell-Yan kinematic configuration that can mimic a
    lepton scattering process.\label{fig:DDY-danger}}
\end{figure}

NDY pairs can arise in all SM processes where we produce four
leptons. The double Drell-Yan (DDY) process $q{\bar q} \to
l^+l^-\ell^+\ell^-$ represents a particularly serious background
to the lepton initiated NDY pair creation. 
First of all, the process has a double collinear enhancement when the
opposite-sign, same-flavour pairs have a small mass. Especially at the
low transverse momenta that are required to get a reasonable yield at
the LHC, there is a substantial phase space for two of the leptons to
be produced below a detection threshold. If, for example, the $l^-$
and $\ell^-$ leptons have transverse momenta below a GeV, they will
escape detection, and the signal will look like a pair of nearly
balanced different-flavour leptons, as depicted in
Fig.~\ref{fig:DDY-danger}.
Thus, when setting up our cuts, we will also consider their effect on the
DDY background.

\begin{figure}[htb]
  \resizebox{7cm}{!}{\includegraphics{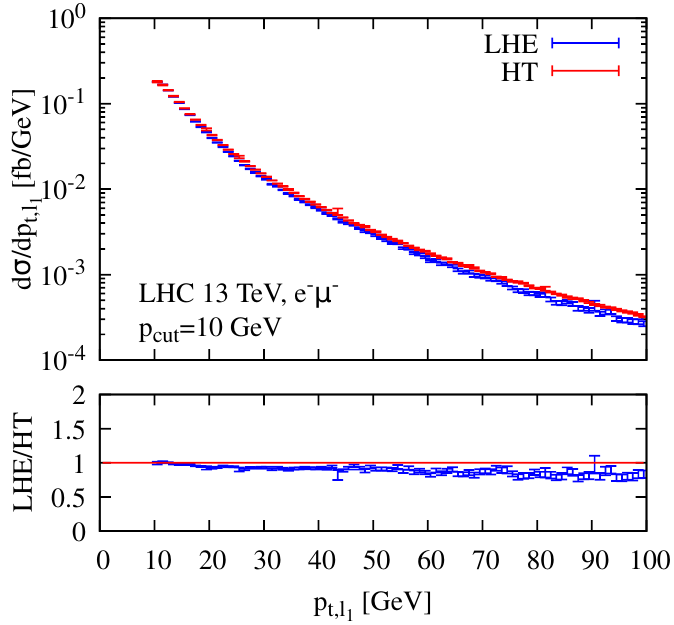}}
  \resizebox{7.3cm}{!}{\includegraphics{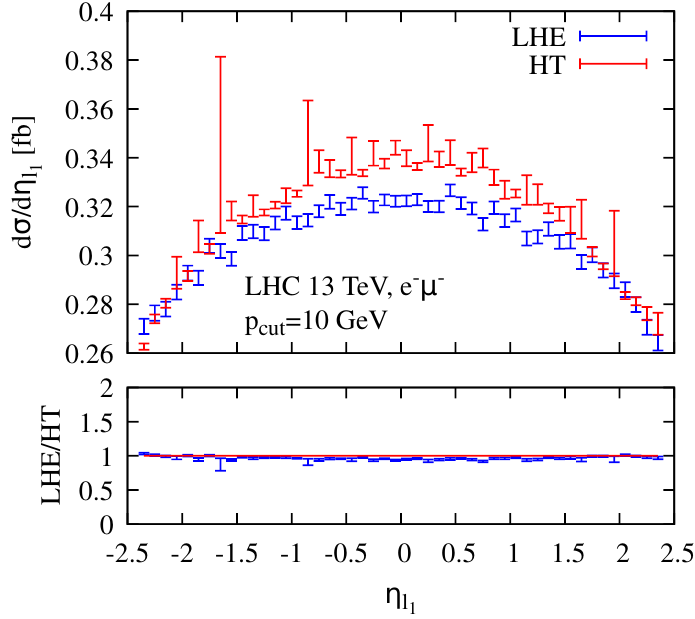}}
  \caption{\label{fig:NLO-SF-pt-eta} Transverse momentum and rapidity
    of the electron, for the case of $e^-\mu^-$ scattering, according
    to the \LHENLO{} result (blue) and the \HT{} calculation (red) for
    $p_{\tmop{cut}} = 10$~GeV.}
\end{figure}

\begin{figure}[htb]
  \resizebox{7cm}{!}{\includegraphics{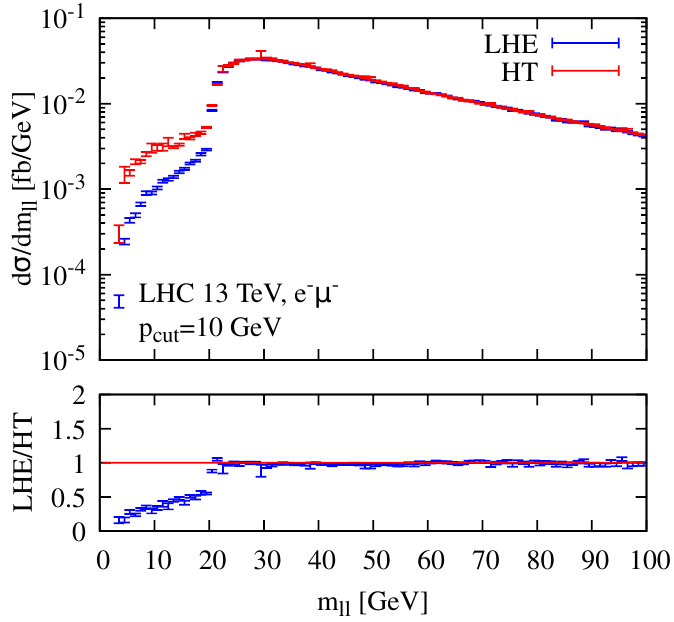}}
  \resizebox{7cm}{!}{\includegraphics{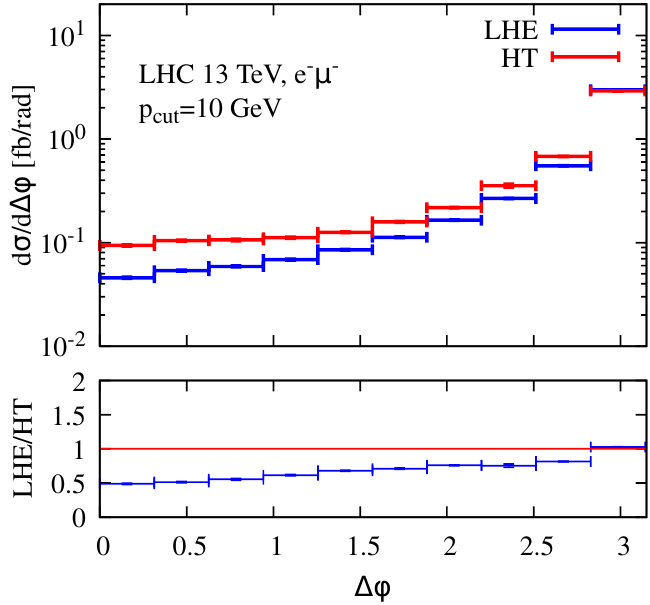}}
  \caption{\label{fig:NLO-SF-mll-azi} Invariant mass and azimuthal
    difference of the signal leptons, for the case of $e^-\mu^-$
    scattering, according to the \LHENLO{} result (blue) and the \HT{}
    calculation (red) for $p_{\tmop{cut}} = 10$~GeV.}
\end{figure}

\begin{figure}[htb]
\centering
\includegraphics[width=0.40\textwidth]{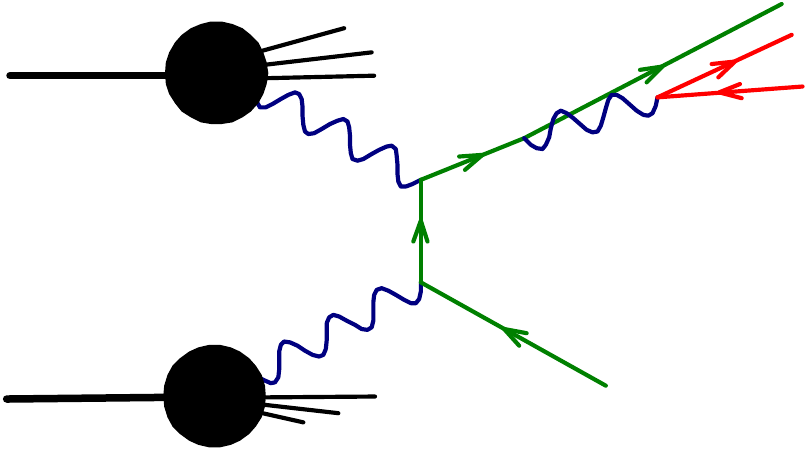}
\caption{\label{fig:lowmasslpair} Configurations that contribute to an
  enhanced production of low mass signal lepton pairs having large
  transverse momenta in the \HT{} approach.}
\end{figure}

\begin{figure}[htb]
  \resizebox{7cm}{!}{\includegraphics{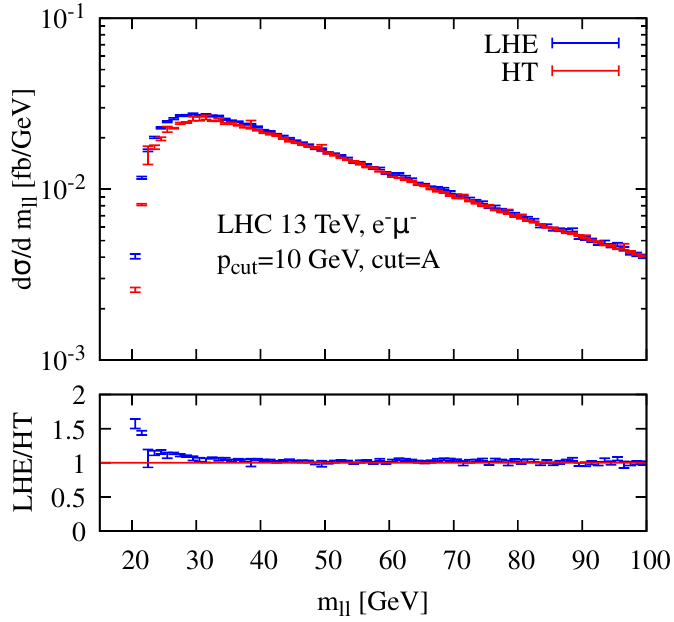}}
  \resizebox{7cm}{!}{\includegraphics{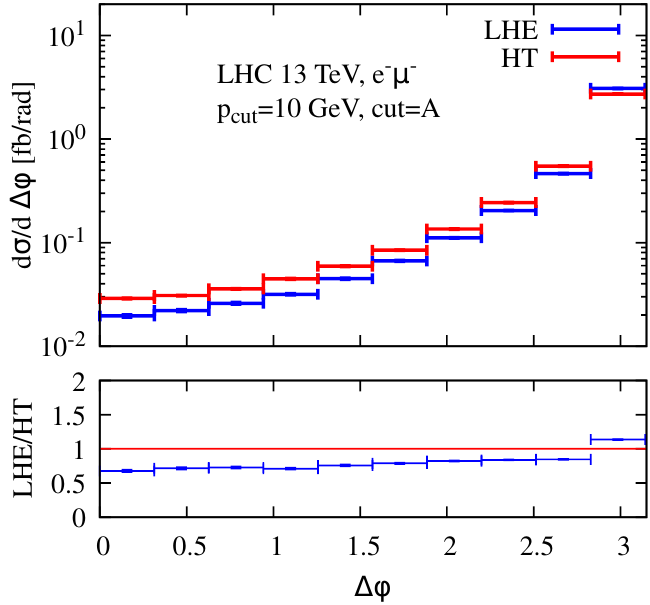}}
  \caption{\label{fig:NLO-SF-mll-azi-cut-A}
  As in figure \ref{fig:NLO-SF-mll-azi}, after imposing cut A.}
\end{figure}

\begin{figure}[htb]
  \resizebox{7cm}{!}{\includegraphics{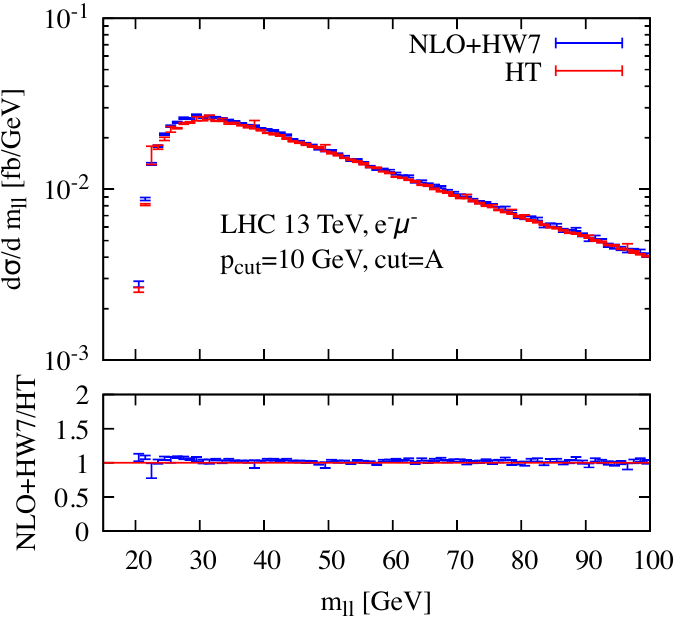}}
  \resizebox{7cm}{!}{\includegraphics{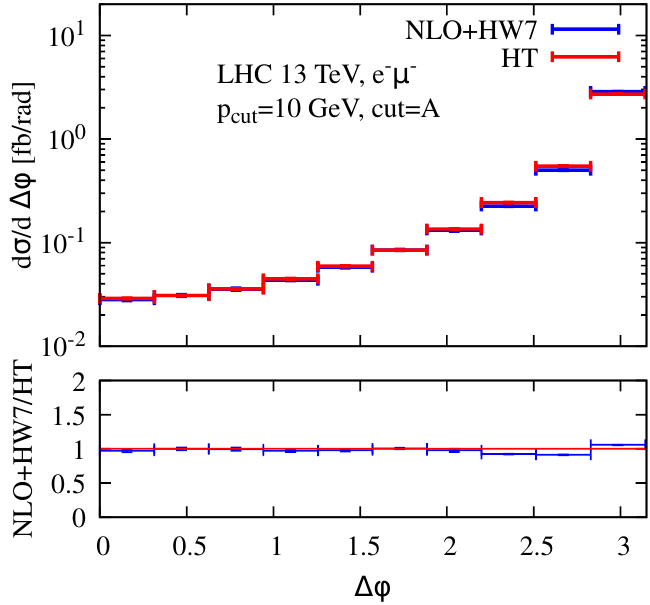}}
  \caption{\label{fig:HW7-SF-mll-azi-cut-A} As in figure
    \ref{fig:NLO-SF-mll-azi}, after imposing cut A and after full
    shower of the \POWHEG{} result.}
\end{figure}

We begin by showing in Figs.~\ref{fig:NLO-SF-pt-eta} and
\ref{fig:NLO-SF-mll-azi} some differential distributions obtained with
the basic cuts of eq.~(\ref{eq:cuts}) for $e^-\mu^-$ scattering,
comparing the result of the NLO \POWHEG{} generator at the parton
level, i.e. at the level of the Les Houches events file (\LHENLO{}
from now on), with that obtained with the \HT{} method. The most
remarkable difference shows up in the distributions of the invariant
mass of the pair and their azimuthal distance, in
Fig.~\ref{fig:NLO-SF-mll-azi}.  From the figure, it is apparent that,
in the \HT{} calculation, there is a large contribution in the region
where the invariant mass of the $e^-\mu^-$ pair is below twice the
transverse momentum cut. The origin of this contribution is easily
identified as the production of a back-to-back lepton-antilepton pair
with relatively large transverse momentum, with the lepton
subsequently radiating a photon, that in turn splits into a second
lepton-antilepton pair, as depicted in Fig.~\ref{fig:lowmasslpair}.
This mechanism is obviously absent in our NLO+PS calculation, since
the $\gamma\gamma$ initial state for our process arises only at the
next-to-next to leading order (NNLO) level.  The final state has in
this case an $l^- \ell^+ \ell^-$ low mass system at large transverse
momentum, recoiling against a (high transverse momentum) $l^+$ lepton,
and the $l^- \ell^-$ system constitutes our signal. This mechanism
does not operate at the level of the \POWHEG{} generated Les Houches
events, since there no more than three leptons are produced. \ It is
clear that these events are not the ones we are interested in,
especially considering that the DDY process $q \bar{q} \rightarrow l^+
l^- \ell^+ \ell^-$ can generate the same kind of final state, through
the mechanism $q \bar{q} \rightarrow \gamma^{\ast} \rightarrow l^+\,
({l^-}^\ast \rightarrow l^- \,(\gamma^{\ast} \rightarrow \ell^-
\ell^+))$. By requiring $m_{l \ell} > | p_t^{(l)} | + | p_t^{(\ell)}
|$, a condition that we dub ``cut A'' in the following, we strongly
suppress this effect, as shown in Fig.~\ref{fig:NLO-SF-mll-azi-cut-A}.
We see that after cut $A$ is applied, the two calculations are more
consistent.  We notice, however, that the agreement of the two
calculations after cut~A is less than perfect, with the \POWHEG{}
result overshooting the \HT{} one for small invariant mass and
undershooting it for any azimuthal distance away from $\pi$. It is
interesting to see that, after shower, the agreement becomes really
excellent, as shown in Fig.~\ref{fig:HW7-SF-mll-azi-cut-A}.
This result is not too surprising, since the shower Monte Carlo, by adding
also the backward evolution for the incoming quark and photon,
captures the most relevant features already present in the \HT{}
calculation.
In order to reduce the large background from the DDY process $q
\bar{q} \rightarrow l^+ l^- \ell^+ \ell^-$ we also impose the
following cuts
\begin{displaymath}
   | p_t^{l \ell} | / (| p_t^{(l)} | + |
  p_t^{(\ell)} |) < 0.2 \quad \mbox{(cut B) or} \quad 0.1\quad \mbox{(cut C)}\,,
\end{displaymath}
that we have found to be particularly effective.  Thus the full set of
cuts that we have adopted is summarized below:
\begin{enumerateroman}
\item \label{enum:etapt} Both signal leptons must have $|\eta|<2.4$,
  $p_t>p_{\rm cut}$.
\item \label{enum:isol} We veto events with extra leptons with
  $p_t>0.9\,$GeV in a cone of radius $r<0.3$.
\item $m_{l \ell} > | p_t^{(l)} | + | p_t^{(\ell)} |$\quad (cut A)\,,
\item $| p_t^{l \ell} | / (| p_t^{(l)} | + |
  p_t^{(\ell)} |) < 0.2$\quad (cut B) or \quad 0.1\quad (cut C)\,,
\item If there is another lepton passing cuts i and ii with $p_t > 3$ GeV the event is vetoed \quad(cut D).
\end{enumerateroman}
The cuts~(\ref{enum:etapt}) and (\ref{enum:isol}) are always
applied. Their combination will be referred to in the tables
as ``cut T''.

The $e^-\mu^-$ and $e^-\mu^+$
cross sections
obtained by applying various combinations of cuts are reported in
tables~\ref{tab:sigmaWithCuts}
and~\ref{tab:sigmaWithCutsOS}. Following our findings on the scale
dependence in Section~\ref{sec:lepscatNLO} we choose as factorization
scale the sum of the transverse momentum of all final-state leptons
divided by two, ignoring the slight preference for smaller (larger)
scale of the $e^-\mu^-$ ($e^-\mu^+$) processes.
\begin{table}[htb]
  \centering
  \resizebox{\columnwidth}{!}{
\begin{tabular}{|l|c|c|c|c|c|c|}
  \hline
  $e^- \mu^-$ &  T  &  TA  &  TAB & TABD & TAC  & TACD             \\
  \hline
 \multicolumn{7}{|l|}{$\sigma\;\mbox{(fb)},\; p_{\rm cut}=5\,$GeV}  \\
\hline
 LO & $ 5.789^{+ 3.343}_{- 2.267}$ & $ 5.789^{+ 3.343}_{- 2.267}$ & $ 5.789^{+ 3.343}_{- 2.267}$ & $ 5.789^{+ 3.343}_{- 2.267}$ & $ 5.789^{+ 3.343}_{- 2.267}$ & $ 5.789^{+ 3.343}_{- 2.267}$ \\ 
 \hline
 NLO & $ 5.53_{- 0.57}$ & $ 4.41_{- 1.07}$ & $ 2.21^{+ 0.59}_{- 1.95}$ & $ 1.88^{+ 0.69}_{- 2.08}$ & $ 0.5^{+ 1.2}_{- 2.7}$ & $ 0.4^{+ 1.2}_{- 2.7}$ \\ 
 \hline
 \LHENLO{} & $ 6.300_{- 0.762}$ & $ 5.493_{- 0.591}$ & $ 3.802_{- 0.519}$ & $ 3.580_{- 0.544}$ & $ 2.799_{- 0.474}$ & $ 2.764_{- 0.476}$ \\ 
 \hline
 NLO+HW7 & $ 6.646$ & $ 5.532$ & $ 3.255$ & $ 2.858$ & $ 1.934$ & $ 1.804$ \\ 
 \hline
 HT & $ 6.65$ & $ 5.232$ & $ 3.13$ & $ 2.82$ & $ 2.06$ & $ 1.98$ \\ 
 \hline
 DDY & $  104.$ & $  28.5$ & $ 4.33$ & $ 1.22$ & $ 1.35$ & $ 0.42$ \\ 
 \hline
 \multicolumn{7}{|l|}{$\sigma\;\mbox{(fb)},\; p_{\rm cut}=10\,$GeV}  \\
\hline
 LO & $ 1.432^{+ 0.734}_{- 0.520}$ & $ 1.432^{+ 0.734}_{- 0.520}$ & $ 1.432^{+ 0.734}_{- 0.520}$ & $ 1.432^{+ 0.734}_{- 0.520}$ & $ 1.432^{+ 0.734}_{- 0.520}$ & $ 1.432^{+ 0.734}_{- 0.520}$ \\ 
 \hline
 NLO & $ 1.28_{- 0.14}$ & $ 1.03^{+ 0.02}_{- 0.24}$ & $ 0.56^{+ 0.15}_{- 0.40}$ & $ 0.31^{+ 0.23}_{- 0.49}$ & $ 0.2^{+ 0.3}_{- 0.5}$ & $ 0.1^{+ 0.3}_{- 0.6}$ \\ 
 \hline
 \LHENLO{} & $ 1.469_{- 0.128}$ & $ 1.281_{- 0.093}$ & $ 0.920_{- 0.129}$ & $ 0.752_{- 0.145}$ & $ 0.687_{- 0.119}$ & $ 0.652_{- 0.121}$ \\ 
 \hline
 NLO+HW7 & $ 1.488$ & $ 1.262$ & $ 0.847$ & $ 0.664$ & $ 0.563$ & $ 0.496$ \\ 
 \hline
 HT & $ 1.53$ & $ 1.234$ & $ 0.80$ & $ 0.63$ & $ 0.55$ & $ 0.50$ \\ 
 \hline
 DDY & $  51.7$ & $  17.$ & $ 3.02$ & $ 0.47$ & $ 0.95$ & $ 0.2$ \\ 
 \hline
 \multicolumn{7}{|l|}{$\sigma\;\mbox{(fb)},\; p_{\rm cut}=20\,$GeV}  \\
\hline
 LO & $ 0.315^{+ 0.143}_{- 0.105}$ & $ 0.315^{+ 0.143}_{- 0.105}$ & $ 0.315^{+ 0.143}_{- 0.105}$ & $ 0.315^{+ 0.143}_{- 0.105}$ & $ 0.315^{+ 0.143}_{- 0.105}$ & $ 0.315^{+ 0.143}_{- 0.105}$ \\ 
 \hline
 NLO & $ 0.28_{- 0.03}$ & $ 0.22^{+ 0.01}_{- 0.05}$ & $ 0.13^{+ 0.03}_{- 0.08}$ & $0.397^{+0.563}_{-1.086} \times 10^{ -1}$ & $0.570^{+0.502}_{-1.023} \times 10^{ -1}$ & $0.97^{+6.32}_{-11.81} \times 10^{ -2}$ \\ 
 \hline
 \LHENLO{} & $ 0.313_{- 0.019}$ & $ 0.270_{- 0.016}$ & $ 0.199_{- 0.027}$ & $ 0.14^{+ 0.01}_{- 0.03}$ & $ 0.15_{- 0.03}$ & $ 0.12_{- 0.03}$ \\ 
 \hline
 NLO+HW7 & $ 0.309$ & $ 0.262$ & $ 0.186$ & $ 0.12$ & $ 0.14$ & $ 0.11$ \\ 
 \hline
 HT & $ 0.322$ & $ 0.261$ & $ 0.179$ & $ 0.122$ & $ 0.129$ & $ 0.104$ \\ 
 \hline
 DDY & $  20.$ & $ 8.26$ & $ 2.18$ & $ 0.2$ & $ 0.72$ & $0.99 \times 10^{ -1}$ \\ 
 \hline
\end{tabular}}
\caption{\label{tab:sigmaWithCuts} Cross sections in femtobarns for
  the production of a $e^-\mu^-$ pair with several combinations of cuts, as
  detailed in the text, in 13~TeV $pp$ collisions, obtained with:
  leading order calculation with lepton parton densities; NLO fixed
  order calculation with lepton parton densities; NLO cross section as
  generated with \POWHEG{} at the Les Houches event level; full
  \POWHEG{} result after shower with \Herwig{}; the \HT{}; and the double
  Drell-Yan four lepton production $q{\bar q}\to \mu \mu^+ e e^+$
  computed at leading order.
  For the LO, NLO and LHE results the quoted uncertainty correspond to the
  three-point scale variation around the central scale choice.}
\end{table}
\begin{table}[htb]
  \centering
  \resizebox{\columnwidth}{!}{
\begin{tabular}{|l|c|c|c|c|c|c|}
  \hline
  $e^-\mu^+$ &  T  &  TA  &  TAB & TABD & TAC  & TACD             \\
  \hline
 \multicolumn{7}{|l|}{$\sigma\;\mbox{(fb)},\; p_{\rm cut}=5\,$GeV}  \\
\hline
 LO & $ 5.737^{+ 3.316}_{- 2.248}$ & $ 5.737^{+ 3.316}_{- 2.248}$ & $ 5.737^{+ 3.316}_{- 2.248}$ & $ 5.737^{+ 3.316}_{- 2.248}$ & $ 5.737^{+ 3.316}_{- 2.248}$ & $ 5.737^{+ 3.316}_{- 2.248}$ \\ 
 \hline
 NLO & $ 6.82^{+ 0.10}_{- 0.90}$ & $ 5.70_{- 0.54}$ & $ 3.35^{+ 0.18}_{- 1.35}$ & $ 2.95^{+ 0.31}_{- 1.52}$ & $ 1.36^{+ 0.84}_{- 2.24}$ & $ 1.29^{+ 0.86}_{- 2.26}$ \\ 
 \hline
 \LHENLO{} & $ 7.422^{+ 0.603}_{- 1.142}$ & $ 6.595^{+ 0.424}_{- 0.957}$ & $ 4.753^{+ 0.067}_{- 0.572}$ & $ 4.459^{+ 0.021}_{- 0.518}$ & $ 3.483_{- 0.395}$ & $ 3.437_{- 0.388}$ \\ 
 \hline
 NLO+HW7 & $ 7.79$ & $ 6.65$ & $ 4.00$ & $ 3.42$ & $ 2.295$ & $ 2.098$ \\ 
 \hline
 HT & $ 7.41$ & $ 5.84$ & $ 3.57$ & $ 3.18$ & $ 2.30$ & $ 2.20$ \\ 
 \hline
 DDY & $  105.$ & $  28.8$ & $ 4.55$ & $ 1.29$ & $ 1.41$ & $ 0.45$ \\ 
 \hline
 \multicolumn{7}{|l|}{$\sigma\;\mbox{(fb)},\; p_{\rm cut}=10\,$GeV}  \\
\hline
 LO & $ 1.396^{+ 0.717}_{- 0.507}$ & $ 1.396^{+ 0.717}_{- 0.507}$ & $ 1.396^{+ 0.717}_{- 0.507}$ & $ 1.396^{+ 0.717}_{- 0.507}$ & $ 1.396^{+ 0.717}_{- 0.507}$ & $ 1.396^{+ 0.717}_{- 0.507}$ \\ 
 \hline
 NLO & $ 1.56_{- 0.15}$ & $ 1.31_{- 0.10}$ & $ 0.81^{+ 0.07}_{- 0.27}$ & $ 0.51^{+ 0.16}_{- 0.39}$ & $ 0.37^{+ 0.20}_{- 0.45}$ & $ 0.30^{+ 0.22}_{- 0.47}$ \\ 
 \hline
 \LHENLO{} & $ 1.720^{+ 0.081}_{- 0.216}$ & $ 1.526^{+ 0.046}_{- 0.178}$ & $ 1.130_{- 0.105}$ & $ 0.907_{- 0.066}$ & $ 0.837_{- 0.070}$ & $ 0.790_{- 0.063}$ \\ 
 \hline
 NLO+HW7 & $ 1.76$ & $ 1.53$ & $ 1.05$ & $ 0.78$ & $ 0.67$ & $ 0.57$ \\ 
 \hline
 HT & $ 1.67$ & $ 1.365$ & $ 0.895$ & $ 0.680$ & $ 0.600$ & $ 0.537$ \\ 
 \hline
 DDY & $  52.3$ & $  17.$ & $ 3.21$ & $ 0.51$ & $ 1.01$ & $ 0.2$ \\ 
 \hline
 \multicolumn{7}{|l|}{$\sigma\;\mbox{(fb)},\; p_{\rm cut}=20\,$GeV}  \\
\hline
 LO & $ 0.294^{+ 0.133}_{- 0.099}$ & $ 0.294^{+ 0.133}_{- 0.099}$ & $ 0.294^{+ 0.133}_{- 0.099}$ & $ 0.294^{+ 0.133}_{- 0.099}$ & $ 0.294^{+ 0.133}_{- 0.099}$ & $ 0.294^{+ 0.133}_{- 0.099}$ \\ 
 \hline
 NLO & $ 0.33_{- 0.03}$ & $ 0.27_{- 0.03}$ & $ 0.18^{+ 0.01}_{- 0.05}$ & $0.674^{+0.437}_{-0.909} \times 10^{ -1}$ & $0.942^{+0.351}_{-0.813} \times 10^{ -1}$ & $0.385^{+0.504}_{-1.001} \times 10^{ -1}$ \\ 
 \hline
 \LHENLO{} & $ 0.359^{+ 0.013}_{- 0.037}$ & $ 0.315^{+ 0.006}_{- 0.029}$ & $ 0.236_{- 0.016}$ & $ 0.16_{- 0.01}$ & $ 0.177_{- 0.009}$ & $ 0.14_{- 0.01}$ \\ 
 \hline
 NLO+HW7 & $ 0.37$ & $ 0.32$ & $ 0.22$ & $ 0.14$ & $ 0.16$ & $ 0.12$ \\ 
 \hline
 HT & $ 0.35$ & $ 0.29$ & $ 0.20$ & $ 0.13$ & $ 0.14$ & $ 0.11$ \\ 
 \hline
 DDY & $  21.$ & $ 8.44$ & $ 2.31$ & $ 0.3$ & $ 0.77$ & $ 0.1$ \\ 
 \hline
\end{tabular}}
\caption{\label{tab:sigmaWithCutsOS}
 As in table~\ref{tab:sigmaWithCuts} for the $e^- \mu^+$ process.}
\end{table}
First of all, we notice that the NLO+PS and \HT{} results are in good
agreement for all combinations of cuts that we considered.  We also see
that when the simplest cuts are applied, i.e. cut T, all results are
in reasonable agreement. On the other hand, as the complexity of the
applied cuts increases, the fixed order results departs from the
LH-NLO and NLO+PS ones.
This is not surprising, and it is a recurring characteristic of fixed
order NLO calculations when the events are cut in order to make them
more similar to the Born ones.  For example, in our case, when we
impose the $p_t$ balance with cuts B and C, we reduce the size of the
real contributions.  These are in fact divergent, and are rendered
finite by the inclusion of virtual corrections.  Asking eventually
perfect balance would veto all real events, yielding an infinite
negative result.  Thus, depending upon the cuts, one can get results
that are as small as one pleases, and that can even become negative at
fixed order. This effect is quite visible in our table, where, as the
cuts become more aggressive, the NLO results decrease and their scale
dependence becomes larger, yielding also negative values.

\begin{figure}[htb]
  \centering
  \includegraphics[width=0.49\textwidth]{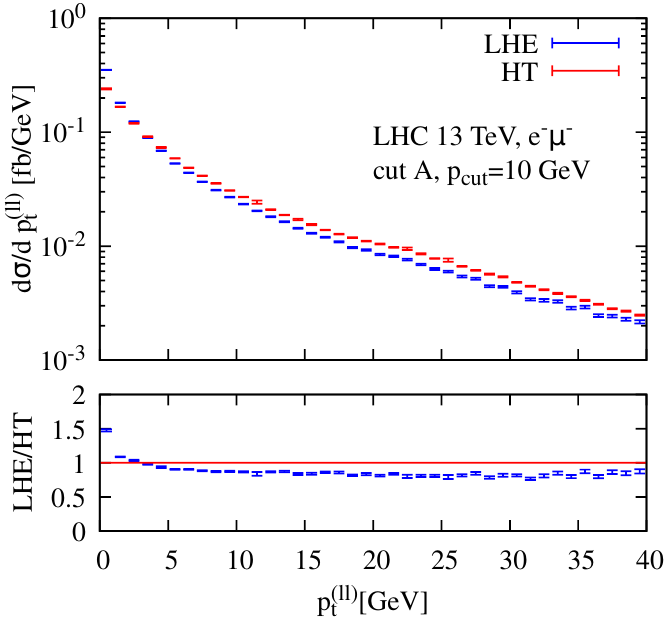}
  \includegraphics[width=0.49\textwidth]{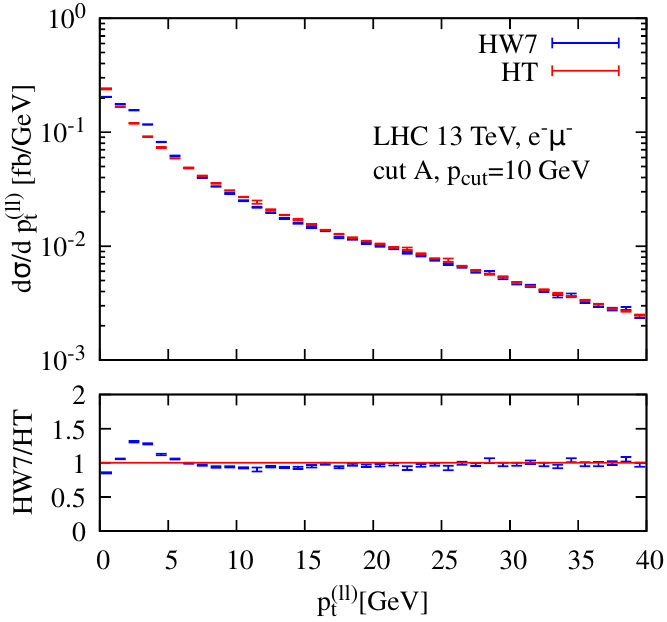}
  \caption{\label{fig:ptllLHHW} The transverse momentum spectrum of
    the two-leptons system defined by the two signal leptons, as
    computed with \POWHEG{} before showering (labeled LHE) (on the
    left), and after the inclusion of the \Herwig{} shower (on the
    right), both compared to the \HT{} result. }
  \end{figure}

We stress that the LH-NLO result by itself is actually incomplete.
For example the transverse momentum of the three hardest leptons is
strictly zero at the LH-NLO level. We expect that backward evolution
in the shower should lift this constraint, and yield a reasonable
transverse momentum distribution for this system, and also reduce
further the number of balanced events (i.e. those that pass cut B and
C). Therefore we expect that after
shower the results should be in better agreement with the \HT{}
calculation, that has always the incoming leptons resolved into
splitting photons, with the photons in turn being produced in
association with a jet (that in the elastic case may consist of a
single proton), and  in fact this is what we observe in our table.
We also show in Fig.~\ref{fig:ptllLHHW} the transverse
momentum of the signal-leptons pair before and after shower compared
with the \HT{} calculation.
  We see the excess in the first bin of the distribution when the
  shower is absent, that is spread out at larger transverse momenta
  when the shower is included, yielding a better agreement with the
  \HT{} result.

As a last comment, we see from the tables~\ref{tab:sigmaWithCuts}
and~\ref{tab:sigmaWithCutsOS} that the combination of cuts $A+B+D$ and
$A+C+D$ are most effective in suppressing the DDY background.  The
effect of the B/C and D cuts is better understood from the following
figures.  In Fig.~\ref{fig:SigVsBg-05-A}, on the left panel, we show
the electron transverse momentum distribution computed at the NLO+PS
level in comparison with the DDY result when only cut A is imposed.
\begin{figure}[htb]
  \resizebox{7cm}{!}{\includegraphics{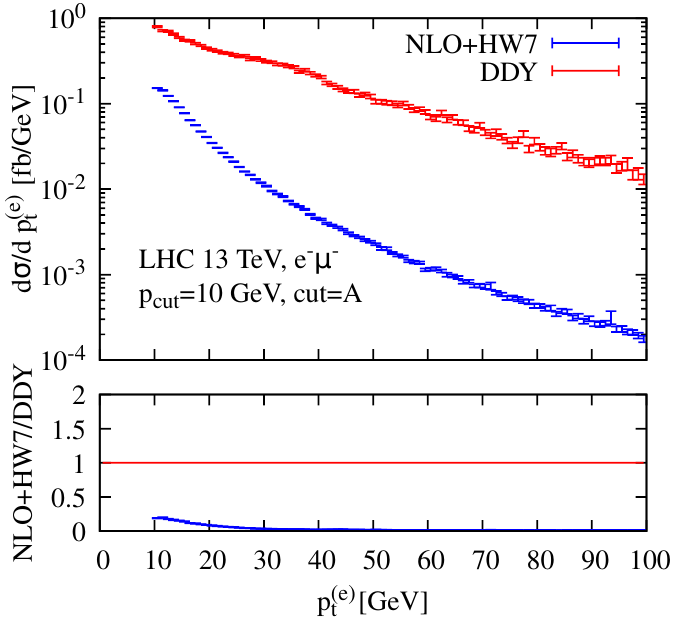}}
  \resizebox{7cm}{!}{\includegraphics{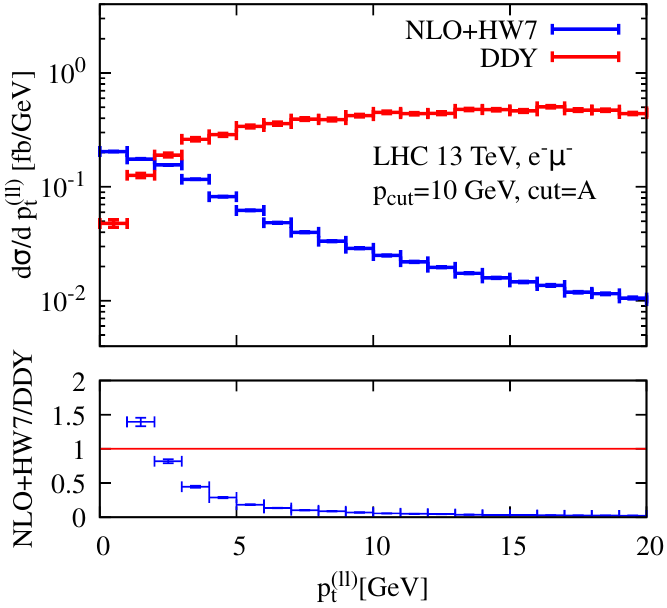}}
  \caption{\label{fig:SigVsBg-05-A} Electron transverse momentum
    (left) and transverse momentum of the pair (right) for the lepton
    scattering signal and the double Drell-Yan background (DDY) when
    cut A is applied.}
\end{figure}
We see that the lepton spectrum is dominated by the DDY process. On
the other hand, we see from the right panel that at low transverse
momenta of the lepton pair, the lepton scattering process prevails.
The same result when applying the A+B cuts is shown in
fig.~\ref{fig:SigVsBg-05-AB}.
\begin{figure}[htb]
  \resizebox{7cm}{!}{\includegraphics{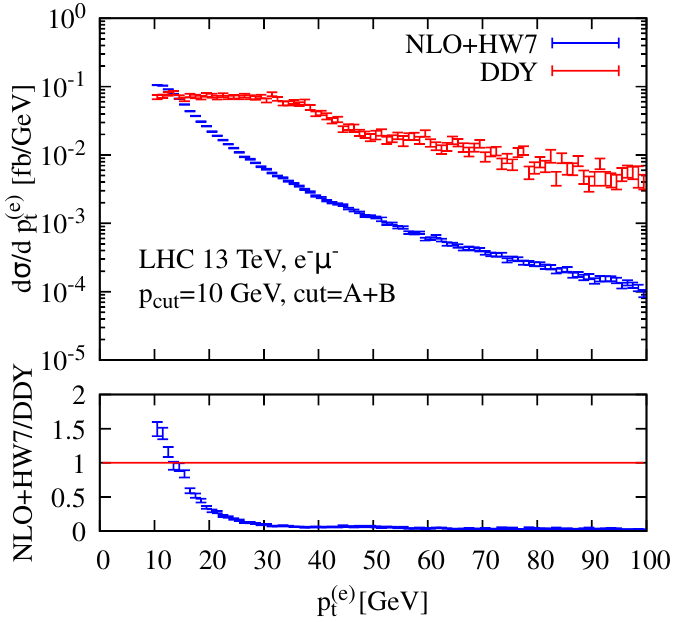}} 
  \resizebox{7cm}{!}{\includegraphics{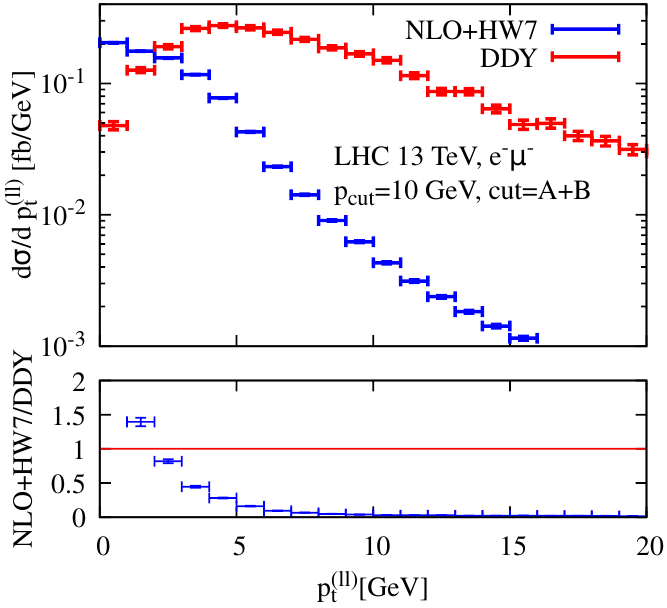}}
  \caption{\label{fig:SigVsBg-05-AB} As in fig.~\ref{fig:SigVsBg-05-A}, for cut A+B.}
\end{figure}
We see that now the signal in the electron $p_t$ emerges from the
background at moderate transverse momenta.  By adding the further
requirement of vetoing other leptons (cut D) we obtain the result
displayed in fig.~\ref{fig:SigVsBg-05-ABD}.
\begin{figure}[htb]
  \resizebox{7cm}{!}{\includegraphics{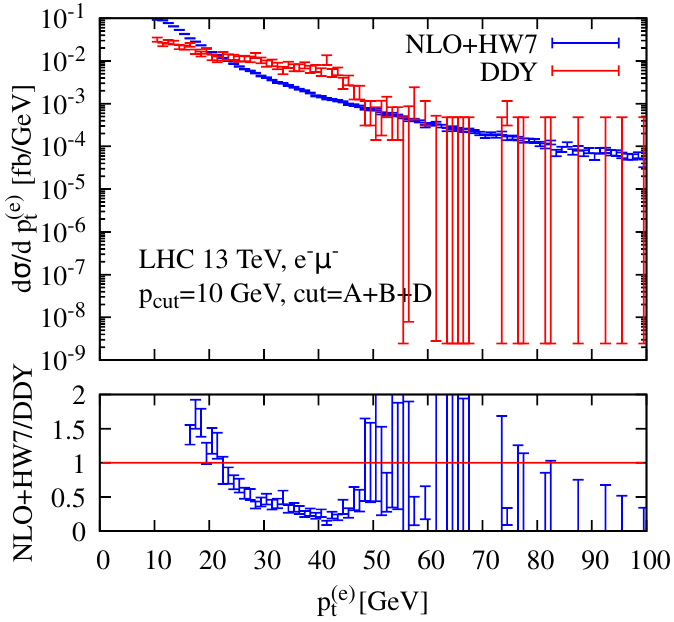}}
  \resizebox{7cm}{!}{\includegraphics{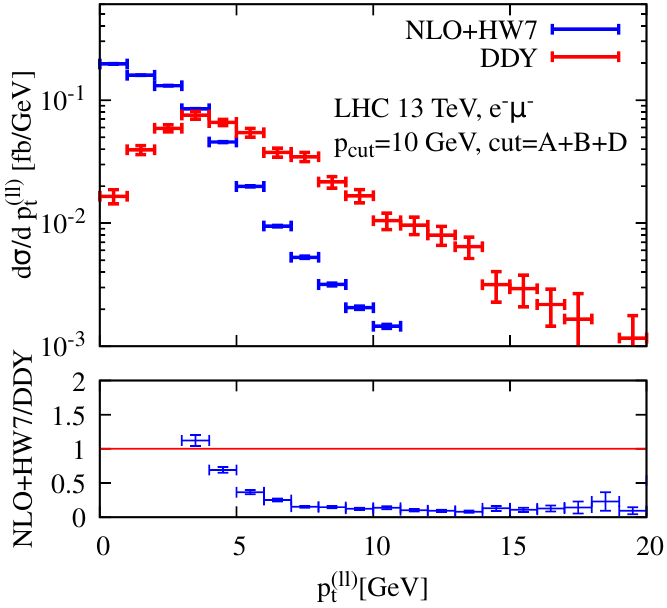}}
  \caption{\label{fig:SigVsBg-05-ABD} As in fig.~\ref{fig:SigVsBg-05-A}, for cut A+B+D.}
\end{figure}
We see that with these cuts the signal prevails over the background as
long as the transverse momentum is not too large. This is due to the
fact that the DDY process is initiated by one valence quark, while the
lepton PDFs are very soft.  Some room is left in tightening the cut on
the balance, going from B to C, in order to further improve the
background rejection, as can be seen in table~\ref{tab:sigmaWithCuts}.

Before ending this section, we turn our attention to
Fig.~\ref{fig:ptllLHHW}, where there is a comparison of the transverse
momentum distribution of the signal pair obtained at the NLO+PS and
\HT{} level.  It is quite clear that the modeling of the first few
bins is less than satisfactory.  We stress that in order to obtain
this level of
agreement
we had to lower the minimum radiation scale of \POWHEG{}
from $0.8$ to $0.1$~GeV, and some parameter tuning on the \Herwig{}
side was also necessary.
This, together with similar observations regarding the
\elgaminit{} scalar production, reminds us that the current status of
the shower generators for \elgaminit{} processes is perhaps not quite
mature, and, as the interest in these processes grows, may require
further work.  

\section{Conclusions}
\label{sec:conclu}

In this paper we have used the recently developed photon and lepton
PDFs to study the impact of NLO corrections to a number of
\elgaminit{} processes and their matching to a parton shower.
While at hadron colliders there is considerable experience in matching
higher-order calculations to parton showers for pure QCD processes,
little is known about this matching in the case of \elgam{}-induced
processes.
For this reason, we have first considered two very simple processes,
namely the production of a
massive scalar particle which couples to incoming photons or leptons.
We have found that LO results display a strong scale dependence,
which translates into very large uncertainty bands, of the order of
${\cal O}(30\%)$ for \gam{}- and ${\cal O}(50\%)$ for \el{}-induced processes.
On the other hand, NLO predictions are stable when varying the
factorization scale, with missing higher-order uncertainties of the
level of 5\% for \gam{}- and 10\% for \el{}-induced production.
It is possible to find a choice for the central scale such that the
NLO corrections are small, ranging from a few percent to about 20\%.
For photon induced processes an optimal choice is the mass of the
scalar $M_\Phi$, while for \elinit{} processes a scale
close to $M_\Phi/4$ is preferred.
We notice in particular that, in the case of \elinit{} processes,
the scale choice $M_\Phi$ gives rise to large and negative
corrections, with the central LO and NLO results differing by a factor
2 and with no overlap between the two uncertainty bands.
Thus limits for new physics based upon searches using
LO predictions at a scale
$M_\Phi$ would be overly aggressive, and it would be better to either
use a lower scale, or rely on a full NLO calculation. 

For photon induced processes we matched the LO and NLO calculations to
both \Pythia{} and \Herwig{} using \POWHEG{}. We find considerable
differences between the two predictions at LO, while we obtain a much
better agreement after matching to NLO.
On the other hand, the two shower Monte
Carlo predictions for the very low transverse momentum distribution
of the scaler have a very different shape. These
features are not relevant for searches of heavy resonances. However,
they give an indication that a more accurate tuning of the shower
generators is desirable for an accurate modeling of \gaminit{}
processes.
We note that, while in general Monte Carlos are tuned to data, in the
present case the tuning could be performed using predictions obtained
with the \HT{} method, since they receive no strong corrections.

The state of the art for \elinit{} processes in shower Monte Carlo programs
is even less advanced
than for \gaminit{} ones, since they are currently implemented only in
a preliminary version of \Herwig{}.
When comparing LO and NLO matched predictions for the transverse momenta
of the scalar and of the accompanying jets/leptons, obtained with the
optimal scale choices ($M_\Phi$ for \gam{}- and $M_\Phi/4$ for
\elinit{} processes), we observe a rather flat ratio as long as the
transverse momenta are not too small. However, we find considerable
differences at small transverse momenta, where NLO corrections have
an opposite effect in \gam{}- and \elinit{} processes.
Concerning the \elinit{} processes we find, as expected, that
\Herwig{} generates small associated hadronic activity, with the
transverse momentum spectrum of the leading jet being comparable to
the one of the sub-leading lepton. This information is important to
enhance the signal to background ratio when looking for these
processes.
Once available also in \Pythia{}, it would be certainly interesting to
compare \Herwig{} and \Pythia{} matched predictions also for \elinit{}
processes.

We have then considered Standard Model $2\to2$ lepton-scattering processes,
focusing our attention on final states which are not Drell-Yan like, and 
extending to NLO accuracy the LO results presented in the \LUXlep{}
paper.
While the NLO calculation is per-se straightforward, care must taken
in its implementation in \POWHEG{}, since, starting from NLO, the
calculation receives contributions from other electromagnetic
sub-processes which involve additional initial- and final-state
collinear singularities.
In particular, because of the latter, at NLO one needs to impose 
additional isolation cuts for the final-state signal leptons.

For the lepton-scattering processes we have also performed a
calculation in the \HT{} approach, which uses the full hadronic tensor
and therefore receives no higher-order QCD corrections. Since the
\HT{} calculation misses instead collinear emissions of
photons from initial-state leptons, we have estimated these effects by
deriving a PDF set where such collinear emissions are switched
off. We find that their
impact is below one percent for the processes at hand.
This implies that one can use the \HT{}
calculation as a reference prediction.  We note however, that the
\HT{} calculation is much more involved than an NLO calculation and
that its matching to a parton shower is not
straightforward.

As in the case of the production of a scalar particle, we find a
considerable reduction in the scale uncertainty when going from LO to
NLO. In general, setting the factorization scale to the transverse
momentum of the outgoing leptons is a reasonable choice,
with slight preferences to lower or higher scales according to the process.

As for the comparison of the \HT{} predictions and the NLO and NLO+PS
calculations, we have found the following features. When minimal
isolation and acceptance cuts are applied, we find discrepancies
between the NLO and the \HT{} result due to a production mechanism
that is included in the \HT{} calculation but not in the NLO one, i.e.
the production of a back-to-back lepton-antilepton pair with
relatively large transverse momentum, followed by the collinear
radiation of a photon that in turn splits into a second collinear
lepton-antilepton pair. This mechanism enters only at NNLO in the
collinear factorization approach.
It is clear that these events are not the ones of interest, in
particular given that the dominant double Drell-Yan (DDY) background process
can generate the same kind of final state. Once one imposes fiducial
cuts to suppress this region one finds good agreement between the
\HT{} and NLO results.

When considering additional cuts, in order to suppress the large DDY
background, the NLO results start departing from the \HT{} one. This
is a typical feature of fixed-order NLO calculations in collider
physics. As one tightens the cuts in order to make the process look
more Born-like, large logarithmically-enhanced negative virtual
corrections are exposed that spoil the reliability of fixed-order
results. However, once the NLO is matched to a parton shower, after a
minimal tuning of the latter, we find very good agreement with the
\HT{} result even after stringent cuts. This is also due to the fact
that the parton shower generates contributions that are present in the
\HT{} calculation and missing in the NLO one.  Furthermore, there is
certainly room for improvements as the implementation of
\elinit{} processes in parton-shower programs is only at its
early stage.


As of now, physics with initial photons and leptons have been mostly
considered for BSM searches. It would be certainly interesting to
experimentally study the lepton-scattering processes, since these are
the only SM processes that can probe the lepton content of the proton.

\section*{Acknowledgments}
We are grateful to Simone Gennai and Michaela Queitsch-Maitland for
fruitful discussions. We thank Peter Richardson for a preliminary
implementation of lepton-induced processes in \Herwig{}, Marco Zaro
for assistance with {\tt MadGradph5} and Silvia Ferrario Ravasio
for assistance with lepton induced processes in
\Herwig{}.
LB is supported by the Swiss National Science Foundation (SNF) under
contract 200020\_188464.  P.N. acknowledges support from Fondazione
Cariplo and Regione Lombardia, grant 2017-2070, and from INFN.

\appendix

\section{The computation of lepton scattering in the hadronic tensor approach}
\label{app:HT}

\begin{figure}[h]
  \centering
  \resizebox{12cm}{!}{\includegraphics{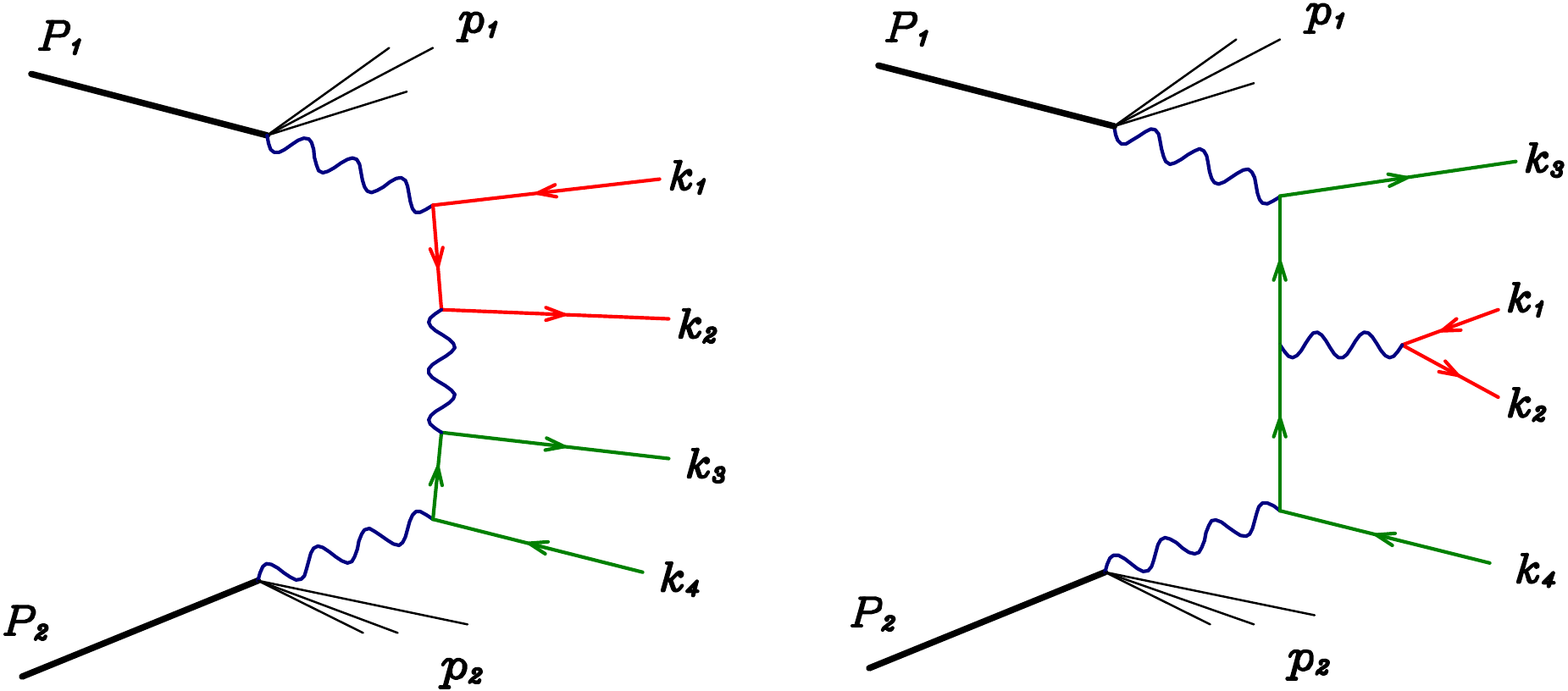}}
  \caption{\label{fig:SF4lep}
    Two sample graphs for the production of two pairs of leptons with
    different flavours. }
\end{figure}

Following the notation of the \LUXlep{} paper, we write the cross section
as
\begin{eqnarray}
  \sigma & = & \frac{1}{2 S}  \int \frac{\mathd^4 q_1}{(2 \pi)^4} 
  \frac{1}{q_1^4}  (4 \pi) W_{\mu \mu'}^{(1)} (P_1, q_1)  \frac{\mathd^4
  q_2}{(2 \pi)^4}  \frac{1}{q_2^4}  (4 \pi) W_{\nu \nu'}^{(2)} (P_2, q_2) \nonumber \\
     &\times&           T^{\mu \mu' ; \nu \nu'}(-q_1,-q_2,k_l)
 \mathd \Gamma (2 \pi)^4 \delta
   \left( - q_1 - q_2 - \sum k_l \right), \label{eq:beginSF}
\end{eqnarray}
where $P_i$ ($i=1,2$) are the momenta of the incoming protons,
$S=(P_1+P_2)^2$, $q_i$ are the four momenta of the virtual photons in
the DIS convention (i.e. directed towards the protons), $k_l$ denote
collectively the momenta of the outgoing leptons, $\mathd \Gamma$ is
the corresponding phase space, and \ $T^{\mu \mu' ; \nu \nu'}$ is the
squared matrix element for the production of four leptons initiated by
the fusion of two off-shell photons.  The process is depicted in
Fig.~\ref{fig:SF4lep},
where we show two sample Feynman diagrams, involving $t$- and
$s$-channel production mechanisms.  We computed the phase space by
composing nested $2\to 2$ scattering processes, loosely following the
approach of Ref.~\cite{Byckling:1969sx}.  We thus rewrite
eq.~(\ref{eq:beginSF}) in terms of the three-momenta $\vec{p}_i$ and
masses $\sqrt{s_i}$ of the outgoing hadronic systems, as
\begin{eqnarray}
  \sigma & = & \frac{1}{2 S}  \int \frac{\mathd s_1}{2 \pi}  \frac{\mathd s_2}{2 \pi} \int
   \frac{\mathd^3 p_1}{2 p_1^0 (2 \pi)^3} \frac{\mathd^3 p_2}{2 p_2^0 (2
               \pi)^3}  \frac{(4 \pi)^2 W_1^{\mu \mu'}(P_1,q_1) W_2^{\nu \nu'}(P_2,q_2)}{Q_1^4 Q_2^4} \nonumber \\
  &\times&  T_{\mu
   \mu' ; \nu \nu'}(-q_1,-q_2,k_l) \mathd \Gamma (2 \pi)^4 \delta \left( P_1 + P_2 - p_1 -
   p_2 - \sum k_l \right),
\end{eqnarray}
where $p^0_i=\sqrt{\vec{p}_i^2+s_i}$, and the $s_i$ must obey the
following constraints
\begin{equation}
 s_i \geqslant m_p^2,\quad\quad \sqrt{s_1} + \sqrt{s_2} + \sum m_l \leqslant \sqrt{S},
\end{equation}
where the $m_l$ are the masses of the outgoing leptons.
The hadronic tensor $W^{\mu \nu}$ is expressed in terms of the
structure functions $F_1$ and $F_2$ (comprising here also
the contribution of the elastic form factors), that in turn are functions of
$Q_i^2=-q_i^2$ and $x_{\tmop{bj},i}$.  The latter is obtained from the
relation
\begin{equation}
s_i = Q_i^2 \frac{1 - x_{\tmop{bj}, i}}{x_{\tmop{bj}, i}} + m_p^2.
\end{equation}
The elastic component of $F_1$ and $F_2$ carry a factor of
\begin{equation}
\delta (1 - x_{\tmop{bj},i}) = Q_i^2\, \delta (s_i - m_p^2).
\end{equation}
We must compute all four combinations of elastic and inelastic
contributions.  Thus, when considering the elastic case, we have to
replace
\begin{equation}
  \int \mathd s_i \rightarrow Q_i^2,
\end{equation}
take the $p_i$ momentum as having mass $m_p$, and write the elastic
$F_i$ contributions without the $\delta(1-x_{\tmop{bj},i})$.

We computed $T_{\mu \mu' ; \nu \nu'}$ using
Madgraph5\_aMC@{}NLO~\cite{Alwall:2014hca} in stand-alone mode. We have
generated the process $\gamma \gamma \rightarrow l^+ l^- \ell^+
\ell^-$ for different and identical fermions. \MADGRAPH{}, by default
generates the diagram for given photon polarizations. It does not
assume, however, that the incoming momenta are on shell. Thus, we
replaced the incoming helicities in the generated formulae with
simple Lorentz indices. More specifically, we implemented a routine 
for the vector-fermion-antifermion vertex that, rather than requiring a helicity in input,
requires a Lorentz index. This routine was obtained
by modifying the \verb!VXXXXX(P,VMASS,NHEL,NSV,VC)! Helas~\cite{Murayama:1992gi} routine
in such a way that, when called with
\verb!NHEL = 0, 1, 2, 3!, it sets all
the \verb!VC(3 : 6)! components to zero except
for the \verb!NHEL+3! one, that is set to
1.

The tensor
$T$ should vanish when contracted with the momenta corresponding with
the two incoming photon, and this served as check for our procedure.

The hadronic tensor is written as
\begin{equation} W_{\mu \mu'} (P, q) = F_1 \left( - g_{\mu \mu'} + \frac{q_{\mu}
   q_{\mu'}}{q^2} \right) + \frac{F_2}{P \cdot q} \left( P_{\mu} - \frac{P
   \cdot q q_{\mu}}{q^2} \right) \left( P_{\mu'} - \frac{P \cdot q
   q_{\mu'}}{q^2} \right), 
\end{equation}
and
\begin{equation} F_L = \left( 1 + \frac{4 x_{\tmop{bj}}^2 m_p^2}{Q^2} \right) F_2 - 2
   x_{\tmop{bj}} F_1, 
\end{equation}
where we have defined $Q^2 = - q^2$. The fitted/calculated values of
$F_2$ and $F_L$ are available in appropriate routines that were
developed in the \LUX{} papers.  Their value is in part extracted from
electro-production data, and in part is calculated in a way that is
described there in great detail.

The phase space in eq.~(\ref{eq:beginSF}) requires great care, since
it is logarithmically singular in several regions, the logarithmic
singularities being regulated by the proton and the lepton
masses. Following Ref.~\cite{Byckling:1969sx}, we factorize the phase
space as a sequence of nested two-body components. Our main
integration variables are thus the $s$-channel and $t$-channel
invariants of these two-body components, that are chosen in such a way
that the logarithmically enhanced regions are adequately sampled.

\subsection{Basic $2 \rightarrow 2$ formulae}
\label{app:basic}
The basic $p_1\,p_2 \rightarrow p_3\,p_4$ cross section formula for
generic masses is
\begin{eqnarray*}
  \mathd \sigma & = & \frac{| \mathcal{M} |^2}{4 \sqrt{(p_1 \cdot p_2)^2 -
  m_1^2 m_2^2}} \mathd \Phi_2\,,\\
  \mathd \Phi_2 & = & (2 \pi)^4 \delta^4 (q - p_3 - p_4)  \frac{\mathd^3
  p_3}{2 p_3^0 (2 \pi)^3}  \frac{\mathd^4 p_4}{(2 \pi)^4} 2 \pi \delta (p_4^2
  - m_4^2)\\
  & = & \frac{1}{32 \pi^2}  \frac{2 \underline{p}_3}{q} \mathd \cos \theta\,,
  \mathd \phi,
\end{eqnarray*}
where $q=p_1+p_2$ and $s=q^2$. We denote as $\underline{p}=\sqrt{\vec{p}^2}$, that in the
above formula is taken in the CM system of $q$.
We have
\begin{equation} p_3^0 = \frac{s + m_3^2 - m_4^2}{2 \sqrt{s}}\,.
\end{equation}
We can introduce
\begin{equation} t = (p_1 - p_3)^2 = m_1^2 + m_3^2 - 2 p_1^0 p_3^0 + 2 \underline{p}_1
   \underline{p}_3 \cos \theta\,, 
\end{equation}
and
\begin{equation} u = (p_2 - p_3)^2 = m_2^2 + m_3^2 - 2 p^0_2 p_3^0 - 2 \underline{p}_1
   \underline{p}_3 \cos \theta\,, 
\end{equation}
so we can also write
\begin{eqnarray*}
  \mathd \sigma & = & \frac{| \mathcal{M} |^2}{4 \sqrt{(p_1 \cdot p_2)^2 -
  m_1^2 m_2^2}}  \left[ \frac{1}{32 \pi^2}  \frac{2 \underline{p}_3}{q} 
  \frac{1}{2 \underline{p}_1 \underline{p}_3} \mathd t \mathd \phi \right]\\
  & = & \frac{| \mathcal{M} |^2}{4 \sqrt{(p_1 \cdot p_2)^2 - m_1^2 m_2^2}} 
  \left[ \frac{1}{16 \pi}  \frac{1}{\sqrt{s}}  \frac{1}{\underline{p}_1}
  \mathd t \frac{\mathd \phi}{2 \pi} \right]\,.
\end{eqnarray*}
As a further simplification we use the identity
\begin{equation}
  (p_1 \cdot p_2)^2 - m_1^2 m_2^2  =  s\, \underline{p}_1^2,
\end{equation}
so we obtain finally
\begin{equation} \mathd \sigma = \frac{| \mathcal{M} |^2}{64 \pi}  \frac{1}{s} 
   \frac{1}{\underline{p}_1^2} \mathd t \frac{\mathd \phi}{2 \pi}\,, 
\end{equation}
where the limits for $t$ are
\begin{equation} m_1^2 + m_3^2 - 2 p_1^0 p_3^0 + 2 \underline{p}_1 \underline{p}_3 > t >
   m_1^2 + m_3^2 - 2 p_1^0 p_3^0 - 2 \underline{p}_1 \underline{p}_3\, . 
\end{equation}

\subsection{Composition of the phase space}
\label{app:PS}
We describe here for simplicity the production of two pairs of leptons of different
flavour, as in Fig.~\ref{fig:SF4lep}. We assume that our signal
leptons are those labeled as 2-3. The phase space is decomposed
according to the following sequence:
\begin{eqnarray*}
  P_1 + P_2 & \rightarrow & p_1 + (k_1 + k_2 + k_3 + k_4 +
  p_2),\\
  (P_1 - p_1) + P_2 & \rightarrow & (k_1 + k_2 + k_3 + k_4) +
  p_2,\\
  (P_1 - p_1) + (P_2 - p_2) & \rightarrow & k_1 + (k_2 + k_3 +
  k_4),\\
  (P_1 - p_1 - k_1) + (P_2 - p_2) & \rightarrow & (k_2 + k_3) +
  k_4,\\
  (P_1 - p_1 - k_1) + (P_2 - p_2 - k_4) & \rightarrow & k_2 +
  k_3,
\end{eqnarray*}
where the momenta in bracket denote a massive system, with an
appropriate bound for its mass. For example, for the first line
of the equation, the minimal mass of $p_1$ is $m_p$, and its maximum
is $\sqrt{S}-(m_1+m_2+m_3+m_4 + m_p)$; the minimal mass of the $(k_1 +
k_2 + k_3 + k_4 + p_2)$ system if $m_1+m_2+m_3+m_4 + m_p$, and its
maximum is $\sqrt{S}-m_p$.

The structure functions have an elastic and inelastic component. 
In the elastic case, $p_i$ has fixed invariant mass equal to $m_p$.
Thus, all components, elastic-elastic, inelastic-elastic, elastic-inelastic
and inelastic-inelastic must be generated independently. As a further point,
the procedure outlined above is adequate for the sampling of the $t$-channel
singularities if lepton 1 is produced mainly along the $P_1$ direction, and
lepton 4 along the $P_2$ one, a region that we here dub region $a$. In order
to perform an adequate sampling, we define an enhancement factor that singles
out region $a$
\begin{equation} f_a = \frac{1}{(P_1 - p_1 - k_1)^2 (P_2 - p_2 - k_4)^2 }, 
\end{equation}
and one that singles out the complementary region $b$ where 1 is produced along the
$P_2$ direction and 4 along  $P_1$
\begin{equation} f_b = \frac{1}{(P_1 - p_1 - k_4)^2 (P_2 - p_2 - k_1)^2 } . 
\end{equation}
We then multiply the cross section by the factor
\begin{equation} \frac{f_a}{f_a + f_b}, 
\end{equation}
thus suppressing region $b$. The contribution of region $b$ is
obtained by generating the phase space with the role of the two
flavours exchanged, providing again the similar suppression
factor.

Several other regions of singularity are present, that however do not
contribute significantly to our signal. In order to speed up
convergence for the distributions that we are interested in, we
suppress these regions with adequate suppression factors. The
generated events have this suppression factor removed, so that one
obtains the exact cross section, but with less dense sampling in the
uninteresting regions. These regions are
\begin{itemizedot}
\item The region when the $k_1, k_2$ and/or the $k_3, k_4$ systems
  have small masses. These regions do not generate isolated leptons
  with the appropriate charge. We thus suppress these regions with a
  factor proportional to these masses.
  
\item The region where the signal leptons $k_3$ and $k_4$ are not in a
  reasonable acceptance range. In particular, they must have a
  sizeable transverse momentum, and their pseudorapidity should be
  within acceptance.  The suppression factors that we adopt are
  \begin{equation} \frac{k_{T, 2}^2}{k_{T, 2}^2 + k^2_{T, \tmop{cut}}} \times \left(
      \frac{y_{\tmop{cut}}}{| y_3 | + y_{\tmop{cut}}} \right)^2 \times \left(
      \frac{m_{12}^2}{m_{12}^2 + m^2_{\tmop{cut}}} \right)^2 \times (2
    \rightarrow 3, 1 \rightarrow 4) . 
  \end{equation}
\end{itemizedot}

\section{Mass corrections in the \POWHEG{} computation}
\label{app:sfsplitmass}

\begin{figure}[htb]
  \centering \includegraphics[scale=0.5]{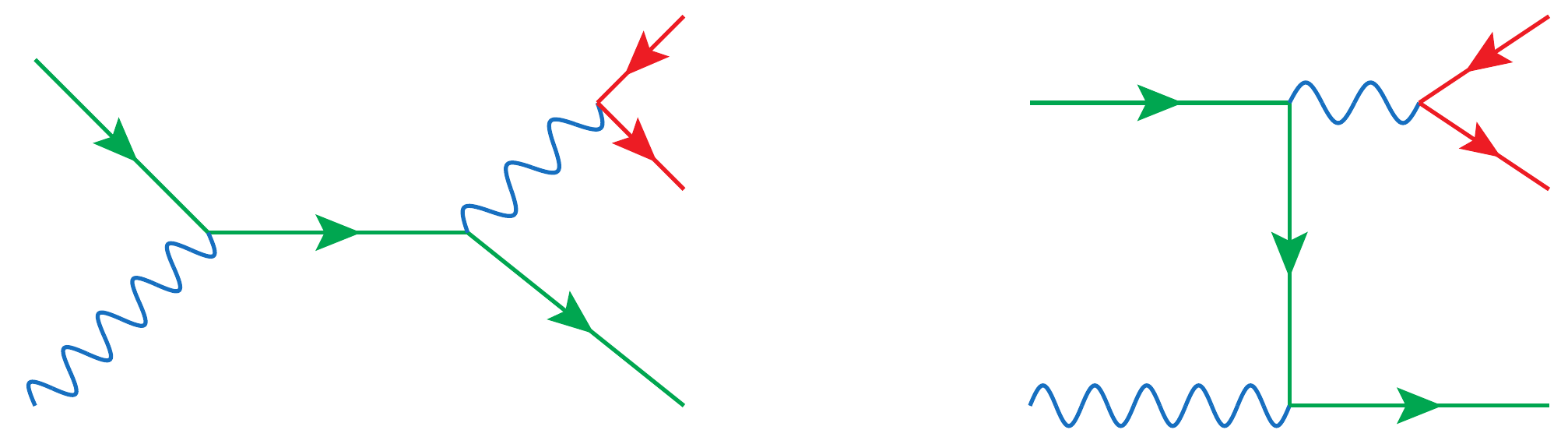}
  \caption{\label{fig:masscorr} Diagrams that lead to a final-state collinear singularity.}
\end{figure}

In this appendix we detail how we include mass effects to cure the
collinear divergence arising from squaring the photon splitting
diagrams depicted in Fig.~\ref{fig:masscorr}. First, let us remark
that this problem does not show up in a complete massive
computation. However, such a computation cannot be performed in the
collinear factorization approach. Second, these
contributions have as underlying Born the Compton process; thus, in
our treatment, they contribute only to the remnant cross section, as
discussed in Sec.~\ref{sec:NLOPWGcalc}. This means that we have to
deal only with the real matrix elements, with no need to modify
subtraction counterterms.

We proceed as follows. We start by computing the square of the two
diagrams in Fig.~\ref{fig:masscorr} assuming the green fermion line,
associated to lepton $l$, as massless and the red one, associated to
lepton $\ell$, as massive. Observe that in the limit for a vanishing
mass the resulting function, that we dub $f_{c}(p;m_{\ell})$, smoothly
approaches the massless squared matrix element that exhibits a
collinear final-state singularity.
Next, we look for all possible $\ell{\bar \ell}$ pairs in the final
state; there are two in the case of identical leptons and just one
otherwise.  For each pair, we let the momenta of these two massless
leptons acquire mass by retaining their 3-momentum while correcting
their energy. The resulting real event has a slightly increased centre
of mass energy. In formulae, writing the original set of momenta $p_i$
in the partonic centre of mass defined by
\begin{equation}
  p_{1,2} = \frac{\sqrt{s}}{2} (1,0,0,\pm1), 
\end{equation}
where $\sqrt{s}$ is the partonic centre of mass energy, the set of
massive momenta is given by
\begin{equation}
  p'_{1,2} = \frac{\sqrt{s'}}{2} (1,0,0,\pm1), \quad p'_{3,4} = \left(
  p_{3,4}^0 \sqrt{1+ \left(\frac{m_{\ell}}{p_{3,4}^0}\right)^2},
  \mathbf{p}_{3,4}\right), \quad p'_{5} = p_{5}
\end{equation}
assuming that the $\ell{\bar \ell}$ pair has momenta $p_{3}$ and
$p_{4}$ and $\sqrt{s'} =
p_{3}^{\prime,0}+p_{4}^{\prime,0}+p_{5}^{\prime,0}$. We correct the
real matrix element by adding the quantity
\begin{equation}
  \Delta_{c} = f_{c}(p';m_{\ell}) - f_{c}(p;0).
\end{equation}
In the case of identical leptons, we add the correction factor
$\Delta_{c}$ for each of the two pairs found, and divide by the
appropriate symmetry factor.

For the sake of completeness, we report here the explicit expression
of the correction function $f_c$ assuming that the collinear leptonic
pair has momenta $p_3$ and $p_4$ and that $p_2$ is the incoming
photon:
\begin{equation}
  f_{c}(p;m_\ell) =-\frac{4}{s_{12}  s_{25} (2 m_\ell^2 + s_{34})^2} (A + 2 B m_{\ell}^2)\,,
\end{equation}
in terms of the invariants $s_{ij} = 2 p_i\cdot p_j$ and the coefficients
\begin{equation}
  \begin{split}
    A &= s_{12} s_{14} s_{23} + s_{14} s_{15} s_{23} + s_{12} s_{13} s_{24} + s_{13} s_{15} s_{24} + s_{12} s_{14} s_{35}  + 2 s_{14} s_{15} s_{35} -  2 s_{13} s_{14} s_{25}  \\
    & + s_{15} s_{24} s_{35} +  s_{14} s_{25} s_{35} + s_{24} s_{25} s_{35} + s_{12} s_{13} s_{45}  +  s_{15} s_{23} s_{45} + 2 s_{13} s_{15} s_{45} - 2 s_{12} s_{35} s_{45}  \\ & + s_{13} s_{25} s_{45} + s_{23} s_{25} s_{45}\,,  \\
    B &= (s_{12}+s_{15})^2 + (s_{15}+s_{25})^2\,.
\end{split}
  \end{equation}

\bibliographystyle{JHEP}
\bibliography{NLO-Lep-Gam}

\end{document}